\documentstyle[pre,aps,eqsecnum]{revtex}
\begin{document}
\title{
Scaling Range and Cutoffs in Empirical Fractals
}
\author{Ofer Malcai\footnote{URL: http://shum.cc.huji.ac.il/~malcai. 
        Email: malcai@flounder.fiz.huji.ac.il},
        Daniel A. Lidar\footnote{Formerly: Hamburger. URL:
        http://www.fh.huji.ac.il/$\sim$dani. Email: dani@batata.fh.huji.ac.il} and
        Ofer Biham\footnote{URL:
        http://www.fiz.huji.ac.il/staff/acc/faculty/biham. Email:
        biham@flounder.fiz.huji.ac.il}} 
\address{
Racah Institute of Physics, The Hebrew University, Jerusalem 91904, Israel
}
\author{David Avnir\footnote{URL: http://chem.ch.huji.ac.il/Avnir.html. 
        Email: avnir@granite.fh.huji.ac.il}}
\address{
Institute of Chemistry and the Minerva Center for 
Computational Quantum Chemistry, The Hebrew University, 
Jerusalem 91904, Israel
}

\maketitle

\begin{abstract}
\newline{}
Fractal structures appear in a vast range of physical systems.
A literature survey 
including 
{\it all experimental papers on fractals} 
which appeared in the six Physical
Review journals 
(A-E and Letters)
during the 1990's shows that
experimental reports of fractal behavior are 
typically based on a scaling range 
$\Delta$
which spans only 0.5 - 2 decades.
This range is limited by upper and lower cutoffs
either because further data is not accessible or due
to crossover bends.
Focusing on spatial fractals,
a classification is proposed into
(a) aggregation; (b) porous media; (c) surfaces and fronts; 
(d) fracture and (e) critical phenomena.
Most of these systems, [except for class (e)]
involve processes far from thermal equilibrium.
The fact that for self similar fractals 
[in contrast to the self affine fractals of class (c)]
there are hardly any exceptions to the finding of 
$\Delta \le 2$ decades, raises the possibility that the
cutoffs are due to intrinsic properties of the measured 
systems rather than the specific 
experimental conditions and apparatus.
To examine the origin of the limited range we
focus on a class of aggregation systems.
In these systems a molecular beam is deposited on a surface, 
giving rise to nucleation and growth of
diffusion-limited-aggregation-like clusters.
Scaling arguments are used to show that the required duration
of the deposition experiment increases exponentially with 
$\Delta$. 
Furthermore, using realistic parameters for surfaces such as
Al(111) it is shown that these considerations limit the
range of fractal behavior to less than two decades in agreement
with the experimental findings.
It is conjectured that related kinetic mechanisms that limit the
scaling range are common in 
other nonequilibrium 
processes which generate
spatial fractals.

\end{abstract}

\pacs{64.60.Ak,61.43.Hv,82.20.Mj,68.55.-a}

\section{Introduction}

The concept of fractal geometry 
\cite{Mandelbrot82,Falconer90}
has proved useful in 
describing structures and processes in experimental systems
\cite{Stanley86,Vicsek89,Feder90,Takayasu90,Avnir92,Bunde94,Barabasi95}.
It provides a framework which can quantify the structural
complexity of a vast range of physical phenomena.
Fractals are objects which exhibit similar
structures over 
a range of length scales
for which
one can define a non-integer dimension.
There are different procedures to evaluate the fractal dimension
of an empirical fractal, all based on multiple resolution analysis.
In this analysis one measures a property $P$ of the system
(such as mass, volume, etc.) as a function of the resolution
used in measuring it (given by a yardstick of linear size $r$).
Fractal objects are characterized by
\begin{equation}
P=k \cdot r^{-D}
\end{equation}
where $D$ is the fractal dimension and $k$ is a prefactor
(related to the lacunarity of the object).
For such objects the graph of $\log P$ vs. $\log r$
exhibits a straight line over a range of length scales
$r_0<r<r_1$
where $r_0$ ($r_1$) is the lower (upper) cutoff. 
The fractal dimension $D$ is given by the slope
of the line within this range.
Typically, the range of linear behavior terminates
on both sides by $r_0$ and $r_1$
either because further data is not accessible
or due to crossover bends
beyond which the slope changes.
For example, in spatial fractals the scaling range is limited 
from below by the 
size of the basic building blocks from which the system is composed
and from above by the system size.
However, 
the empirically measured scaling range may be further reduced 
either due to properties of the measured system or limitations
of the apparatus.
System properties which may further restrict the scaling range
may be: (a) mechanical strength of the object which is reduced
with increasing size; (b) processes which tend to smooth out
the structure and compete with the fractal generating processes;
(c) noise, impurities and other imperfections in the system and
(d) depletion of resources such as space available for growth 
or feed material.
The apparatus may limit the observed scaling range
due to: (a) limited resolution at the smallest scales;
(b) limited scanning area, which may be smaller than the
system size; (c) limited speed of operation which does not
allow to collect enough statistics; (d) constraints in operation
conditions such as temperature, pressure, etc. which may impose
parameters not ideal for the given experiment.

There are different ways to classify empirical fractals.
One classification is according to the type of space in 
which they appear.
This can be: (a) real space; (b) phase space; (c) parameter space
and (d) the time domain (time series).
Spatial fractals
appear in both equilibrium and nonequilibrium systems.
The theory of critical phenomena predicts that at the critical point of 
fluids, magnets and percolation systems the correlation length 
diverges
\cite{Stanley71,Binney92}.
As a result, fractal domain structures appear over all
length-scales up to the system size.
Experimental evidence for fractal structures at criticality has been
obtained for example in the context of percolation
\cite{Kapitulnik84a}, 
in agreement with the theory 
\cite{Kapitulnik84b,Isichenko92}
and computer simulations
\cite{Stauffer79,Stauffer85}.
Reaching the critical point requires fine tuning of
the system parameters, as 
{\it these points are a set of measure zero in parameter space}.
Most empirical fractals have been found in systems far from
thermal equilibrium and thus - not only out of the scope of
critical phenomena, but where equilibrium statistical physics
does not apply.

A variety of dissipative dynamical systems exhibit 
strange attractors with fractal structures in
phase space. 
The theory of dynamical systems provides a 
theoretical framework for the study of fractals in
such systems
at the transition to chaos
and in the chaotic regime
\cite{Eckmann85}.
At the transition to chaos, fractals are found 
also in parameter space 
\cite{Stavans85}
while time series measured in the chaotic regime 
exhibit fractal behavior in the time domain
\cite{Jensen85}.
Fractal dimensions of objects in phase space are not limited by
the space dimension, giving rise to the possibility of $D>3$.
Effective methods for embedding experimental 
time series in higher dimensional
spaces to examine the convergence of fractal dimension calculations 
were developed and widely applied
\cite{Grassberger84}.
However, these should be used with care as the
number of data points
required in order to
measure 
fractal dimensions (FD)
from embedded time series
increases exponentially with
the dimension of the underlying attractor
\cite{Eckmann92}.

In this paper we will focus on 
fractals in real space.
One can classify the spatial fractal structures 
according to physical processes and systems in which they 
appear. We identify the following major classes:
(a) aggregation;
(b) porous media;
(c) surfaces and fronts;
(d) fracture;
(e) critical phenomena (e.g. in magnets, fluids, percolation).
Note that some systems may belong to more than one class.
For example, classes (a) and (d) describe the dynamical processes
which generate the fractal while classes (b) and (c) describe
the structure itself.
Moreover, there is some overlap between (b) and (c) since 
studies of porous media often focus on the fractal structure
of the internal surfaces of the pores
\cite{Pfeifer92}. 
For case (e) of equilibrium critical phenomena there are solid 
theoretical predictions of fractal structures at the critical point,
most extensively examined for the case of percolation
\cite{Kapitulnik84b,Isichenko92}.
The cutoffs in such systems may appear due to small 
deviations of the parameters from the
critical point values
and due to the finite system size.
Spatial fractals in the
four other classes typically result from non-equilibrium 
processes. One
should single out the case of surfaces and 
fronts (c) which are often inherently 
anisotropic and their fractal nature is characterized by self
affine rather than self similar structure
\cite{Barabasi95}. 
Among the other three
classes, 
within the physics literature,
fractals in
aggregation phenomena have been most extensively studied.

The abundance of fractals in aggregation processes stimulated 
much theoretical work in recent years.
The diffusion limited aggregation (DLA) model, introduced
by Witten and Sander
\cite{Witten81,Witten83},
provides much useful insight
into fractal growth
\cite{Meakin88}.
This model includes a single cluster 
to which additional particles attach once they reach a site
adjacent to the edge of the cluster.
The additional particles are 
launched one at a time from random positions far away from the
cluster and move as random walkers until they either attach to
the cluster or move out of the finite system. 
Numerical simulations of this model were used to create very large
fractal clusters 
of up to about 30 million particles
\cite{Mandelbrot92}.
These clusters exhibit
fractal behavior over many orders of magnitude (although
the lacunarity seems to change as a function of the cluster size). 
The asymptotic behavior  
of the DLA cluster has been studied analytically and numerically
for both lattice and continuum models
indicating a considerable degree of universal behavior
\cite{Eckmann90,Arneodo92}.
A universal fractal dimension $D \cong 1.7$ was
observed in two dimensions (2D)
and $D \cong 2.5$ in three dimensions (3D)
\cite{Meakin85}.

Morphologies similar to those of the DLA model
and fractal dimensions around $1.7$ 
have been observed in
a large number of distinct experimental systems. 
These include electrodeposition
\cite{Brady84}
and
molecular beam epitaxy (MBE)
\cite{Hwang91}. 
However, unlike the theoretical model, 
the experimentally observed morphologies are typically
somewhat more compact and the
scaling range does not exceed two orders of magnitude.
This observation has to do with the fact that
unlike theoretical models, 
which may be inherently scale free,
in empirically observed
fractals the range of length-scales over which scaling behavior 
is found is limited by upper and lower cutoffs. 
For finite systems, the scaling range is limited by lower and
upper cutoffs even if the internal structure is scale free.
In this case 
the lower cutoff is the basic unit (or atom) size in the system,
while the upper cutoff is of the order of the system size.
However, typically the scaling range is much narrower than
allowed by the system size, thus limited by other factors.
This width is not predicted by
theoretical models and in many cases not well understood.
There have been some suggestions on how to incorporate
the limited range into the analysis procedure
\cite{Struzik97}.
On the one hand, this range may be simply limited by the apparatus
used in a given experiment. 
If this is the case, 
we would expect to see, at least in some experiments, 
when the most proper apparatus is chosen,
a broad scaling range limited only by the system size.
On the other hand,
the scaling range may be limited 
by properties intrinsic to the system.
In this case, using a different apparatus is not expected to
dramatically broaden the scaling range. 

In this paper we explore the status of experimental 
measurements of fractals. 
Using an extensive survey of experimental  
fractal measurements we examine the 
range of scales in which the fractal behavior is observed and the
fractal dimensions obtained.
We observe a 
a broad distribution of measured dimensions in the range
$0.5<D<3$, most of which 
are interpreted as non universal 
dimensions, that depend on system parameters.
This distribution includes a peak
around $D=1.7$ due to
structures which resemble 2D 
DLA-like clusters, 
which account for a 
significant fraction of the class 
of aggregation processes. 
More importantly,
we find that the range of fractal behavior in experiments
is limited between 0.5 - 2 decades with very few exceptions
as discussed above.
There may be many different reasons for this, which can
be specific to each system or apparatus. 
However, the fact that
the distribution is sharply concentrated around 1.5 decades and
the remarkably small number of exceptions, indicate that there may
be some general common features which limit this range. 
Trying to identify such features,
we focus in this paper
on a class of aggregation problems which appear in MBE experiments.
In these experiments a finite density of DLA-like clusters nucleate
and grow on the substrate. 
The width of the scaling range is limited by the cluster size
(upper cutoff) and the width of its narrow arms (lower cutoff)
which can be as small as the single atom. 
We show that a small increase in the scaling range requires a
large increase in the duration of the MBE experiments.
Moreover, at long times edge diffusion and related processes
which tend to smooth out the fractal structures become 
significant.
These processes tend to increase the lower cutoff
and in this way limit the possibility of further extending 
the scaling range.
This detailed argument is presented only for MBE-like
aggregation problems. 
However, we believe that related arguments, based
on the fact that in empirical systems there is no
complete separation of time-scales, may apply to 
other classes of fractal structures out of equilibrium.

The paper is organized as follows. In Section II we present an
extensive survey of experimental measurements of fractals and
examine the empirical dimensions and scaling range.
In order to obtain better understanding of the limited scaling
range, we focus in Section III on the case of nucleation and
growth of fractal islands on surfaces.
The width of the scaling range is obtained as a 
function of the parameters of the system and it
is shown that under realistic assumptions it does
not exceed two decades.
These results and their implications to empirical 
systems are discussed in Section IV,
followed by a summary in Section V.

\section{Survey of Experimental Results}

Here we present an extensive survey of experimental papers
reporting fractal measurements, and examine the 
range of length-scales
over which fractal properties were observed, 
as well as
the reported dimensions.
In our survey we used the INSPEC data-base from which
we extracted all the {\it experimental} papers in {Physical Review} 
{A - E} and {Physical Review Letters}
over a period of seven years (January 1990 - December 1996)
which include the word {\it fractal} in the title or in the 
abstract, a total of 165 papers
\cite{melvyl}. 
These papers account for $9.1\%$ of the 1821 experimental
papers on fractals that appeared 
during that seven year period 
[and $6.8\%$ of all such papers ever published
(2425 papers since 1978)]
in all scientific journals listed by INSPEC. 

Experimental measurements of fractal dimensions are usually analyzed
using the box counting or related methods. 
In these measurements a log-log plot is reported in which the horizontal
axis represents the length scale (such as the linear box size)
and the vertical axis is some feature
(such as the number of boxes which intersect the fractal set)
for the given box size.
Typically, the reported curves include a range of linear
behavior. 
This range terminates on both sides 
by upper and lower cutoffs
either because
further data is not accessible or due to a knee beyond 
which the line is curved. 
The apparent fractal dimension is then obtained from the slope
of the line in the linear range.
Out of the 165 papers mentioned above, 
86 papers 
\cite{liu90,shih90,fontana90,ausloos90,jian90,devreux90,vacher90
,hwa91,oxaal91,liu91,chachaty91,shang91,birovljev91,constantin91,lemaire91,hwang91,adam91,ferri91
,asnaghi92,robinson92a,robinson92b,mcanulty92,sanchez92,ghosh92,carpineti92,schwartz92
,jensen93,mu93,iwasaki93,zhang93,hasmy93,luo93,sengupta93,buzug93,stankiewicz93,ming93,carro93,wei93,sapoval93,sahimi93,page93
,li94,sen94,xie94,hasmy94,ohsaka94,groote94,ghosh94,shaoshun94,shivpuri94
,vazquez94,wei94,penella94,birovljev94,xiao94,muzny94,holten94,zhu94,cai94,hobbie94,komori94,spector94,kuhn94
,ivanda95,sze95,douketis95,mitropoulos95,shivpuri95,dheer95,abbott95,ghosh95,larsen95,arneodo95,viswanathan95
,wang96,shang96,radlinski96,mattsson96a,mattsson96b,cherry96,du96,balazs96,muller96,bisang96,mallamace96,pignon96}
included such a plot
(and 10 of them included two plots).
For each one of these 96 log-log
plots we extracted both the fractal dimension
and the width of the linear 
range between the cutoffs (Table I)
\cite{linear-log10}.
Table I includes a row for each one of the 96 measurements.
The first column briefly describes the context of 
the experiment. 
The second column provides a classification of the systems
into the following categories:
aggregation (A), 
porous media (P), 
surfaces and fronts (S), 
fracture (F),
critical phenomena (C),
fracton vibrations (V),
turbulence (T),
random walk (R) 
and high energy physics (H).
In cases where more than one class is appropriate we assign
both classes.
The next two columns provide the
fractal dimension (FD) and the width of the scaling range in which fractal
behavior was detected ($\Delta$).
The next three columns provide the lower cutoff
($r_0$), the upper cutoff ($r_1$) and the units in which
these cutoffs are measured.
Note that in many of the papers the scales in the log-log plots are
provided in a dimensionless form or in arbitrary units.
In these cases we left the units column empty.
The last two columns provide 
the Reference number and the
Figure number in that paper from which the FD, $\Delta$ 
and the cutoffs were obtained.
We found that 29 measurements belong to class A, 19 to P,
18 to S, 6 to F, 8 to C, 4 to V, 2 to T, 4 to R and 10 to H.

To examine the distribution of widths of the scaling range
we present a histogram (Fig. 1) which shows,
as a function of the width (in decades)
the number of experimental measurements in which 
a given range of widths was obtained.
Surprisingly,
it is found that the typical range is between 0.5 - 2 decades
with very few exceptions.
To obtain more insight about the scaling range we present
separate histograms for 
aggregation [Fig. 2(a)],
porous media [Fig. 2(b)]
and surfaces and fronts [Fig. 2(c)].
The distribution for aggregation systems is basically similar to
the one of Fig. 1, with a peak around 1.5 decades. 
We note in particular
that it does not include 
measurements over significantly more than two decades.
The width distribution for porous media has the same general shape,
however, the scaling range is typically narrower and 
the peak is centered around one decade.
The width distribution for surfaces and fronts includes
both
a flat range between one and two decades, in addition to 
a few cases with three and four decades. 
It is interesting to note that the papers in which 
three or four decades of scaling behavior are reported 
\cite{birovljev91,sapoval93,birovljev94,holten94}
are in the context of surfaces and fronts, 
related to self affine, rather than self similar fractals.
This observation raises the question whether,
for self similar fractals, there are some
common features of the empirical systems reviewed here, which
tend to limit the width of the scaling range. 

To obtain the distribution of measured fractal dimensions we 
constructed a histogram (Fig. 3) showing the number of experiments 
which observed fractal dimension in a given range. 
The fact that most of the experiments 
deal with spatial fractals
is reflected in the observation that
in most cases $D \le 3$
\cite{Dexceptions}.
Two peaks are identified in the histogram, around
$D \cong 1.7$ and $D \cong 2.5$. 
In addition to these peaks, 
there is a broad distribution
of observed dimensions in the entire range of $0.5<D<3.0$.
To further examine the observed dimensions we also show
separately their distributions for the classes of aggregation 
[Fig. 4(a)], porous media [Fig. 4(b)] and surfaces [Fig. 4(c)].
The statistics available for the other classes is not sufficient
to draw significant conclusions.
We observe that for aggregation systems there is a huge peak
around 
$D \cong 1.7$
which corresponds to 2D DLA.
In addition, there are some systems with higher dimension,
a few of them may correspond to 3D DLA,
for which the dimension is
$D \cong 2.5$.
For porous media we observe a rather flat distribution
of fractal dimensions in the range 
$1.5 < D < 2.8$.
For surfaces and fronts there are two peaks, one around
$D \cong 1.5$
which includes topologically one dimensional fronts and the 
other one around
$D \cong 2.5$
which includes rough two dimensional surfaces.

The measured dimensions in Table I represent not
only empirical measurements of the fractal dimension $D_0$,
but in some cases these are
generalized fractal dimensions. 
In particular, experiments in which scattering
techniques are used tend to provide the correlation dimension
$D_2$. 
The generalized dimension $D_q$ is a monotonically
decreasing function of $q$
\cite{Hentschel83,Halsey86}.

Due to the broad scope of systems included in our survey,
it is not possible at
this stage to provide general arguments.
We chose to focus our discussion on the class of aggregation
systems in which a finite density of DLA-like
clusters nucleate on surfaces.
These systems are in a way representative, 
as they exhibit spatial fractal structures
which grow out of thermal equilibrium.
Moreover, DLA-like
structures account for a significant fraction of the surveyed
papers
and are thus particularly relevant.

\section{DLA-like clusters on surfaces}

We will now examine the scaling properties
and cutoffs in a class of systems
in which DLA-like clusters nucleate and grow
on a surface.
Particularly, in MBE
a beam of atoms is deposited 
on a substrate. 
These atoms diffuse on the surface and nucleate
into islands which keep growing as more atoms are added.
MBE experiments on 
systems such as
Au on Ru(0001)
\cite{Hwang91,Potschke91}, 
Cu on Ru(0001)
\cite{Potschke91}, 
and Pt on Pt(111)
\cite{Bott92,Michely93}
give rise to DLA like clusters
with dimensions close to $1.7$.
We will now consider the growth processes in such experiments.

In MBE experiments atoms are randomly deposited on a clean high
symmetry surface from a beam of flux $F$ 
[given in monolayer (ML) per second]. 
Each atom, upon attachment to
the surface starts hopping as a random walker on a lattice
[which can be a square lattice for FCC(001) substrates
and triangular lattice for FCC(111) substrates]
until it either
nucleates with other atoms to form an immobile cluster or 
joins an existing cluster.
The hopping rate $h_0$ (in unit of hops per second)
for a given atom to each unoccupied  
nearest neighbor site is
\begin{equation}
h_0 = \nu \cdot \exp(-E_0/k_B T)
\label{hoppingrate}
\end{equation}
where $\nu \cong 10^{12}$ is the standardly used attempt frequency,
$E_B$ is the energy barrier, $k_B$ is the Boltzmann
factor and $T$ is the temperature.
The coverage after time $t$ is then
$\theta = F \cdot t$ (in ML).
The submonolayer growth is typically divided into three stages:
the early stage is dominated by island nucleation, followed by
an aggregation dominated stage until coalescence sets in.
In studying the fractal properties of islands we are interested
in the late part of the aggregation stage, where islands are
already large, but separated from each other, 
as coalescence is not yet dominant.
The scaling behavior at this stage has been studied
using both rate equations
\cite{Stoyanov81,Venables84,Villain92,Bartelt92,Tang93,Biham95}
and Monte Carlo (MC) simulations
\cite{Ratsch94,Zhang94,Jacobsen94,Barkema94,Schroeder95,Amar95,Bales95,Linderoth96,Furman97}.
It was found that
the density of islands $N$ is given by
\begin{equation}
N \sim \left( {F \over h_0} \right)^{\gamma}.
\end{equation}
The exponent $\gamma$ is determined by the microscopic processes
that are activated on the surface during growth.
It can be expressed in terms of
the critical island size 
$i^{\ast}$, 
which is 
the size for which all islands 
with a number of atoms 
$s \le i^{\ast}$ 
are unstable (namely dissociate after a short time)
while islands of size 
$s \ge i^{\ast}+1$ are stable.
It was found, 
using scaling arguments and MC simulations 
that for isotropic diffusion,
in the asymptotic limit of slow deposition rate,
$\gamma = i^\ast / (i^\ast + 2)$
\cite{Venables84,Schroeder95,Amar95}.
However, in case that the small islands of size
$s \le i^{\ast}$
are not unstable but only mobile,
the scaling exponent takes the form
$\gamma = i^\ast / (2 i^\ast + 1)$
\cite{Villain92,Furman97}.
For systems in which only the single atom is mobile
(such as the DLA model),
$i^{\ast}=1$ and 
$\gamma = 1/3$
\cite{third}.
The typical distance between the centers of islands,
which is given by
$\ell = N^{-1/2}$
then scales as
\begin{equation}
\ell \sim \left( { h_0 \over F} \right)^{1/6}.
\label{upper_cut}
\end{equation}
The growth potential of
each cluster 
is limited by this distance,
beyond which
it merges with its nearest neighbors.
Therefore, $\ell$ is an upper cutoff
for the scaling range of the DLA-like islands
for the given experimental conditions
\cite{manipulation,Rosenfeld93}.
This cutoff can be pushed up by varying the growth conditions,
namely the temperature and the flux.
However, Eq. 
(\ref{upper_cut}) 
indicates that in order to add one order of
magnitude to $\ell$ one needs to increase the ratio
$h_0/F$ by a factor of $10^6$. 
This can be done either by reducing the
flux, or by raising the temperature, which would increase the
hopping rate.
To get a broad scaling range one can also choose a substrate
with very low hopping barriers, so the required deposition 
rate would not have to be unreasonably small.
However, the slow dependence of $\ell$ on $h_0/F$ indicates
the inherent difficulties in growing fractal islands with a broad
scaling range.

We will now try obtain a more quantitative understanding of the situation.
First, we will consider the case  
of no significant
thickening of the
arms of the DLA-like clusters.
In this case the lower cutoff remains of the order 
of the atom size.
The maximal width of the scaling range,
is then given by 
$\Delta_0 = \log_{10} \ell$,
where $\ell$ is given in units of the substrate lattice 
constant.
We thus obtain:
\begin{equation}
\Delta_0 = {\gamma \over 2} \log_{10} \left( {h_0 \over F} \right).
\label{delta.vs.hf}
\end{equation}
To approach this width the clusters need to fill the
domains of linear size $\ell$  
available to them.
The coverage at which this maximal width is obtained is
\begin{equation}
\theta \sim N \cdot \ell^D \sim 
\left( {F \over h_0} \right)^{\gamma (1- D/2)}, 
\label{maxcov}
\end{equation}
where $D$ is the FD of the clusters
and the deposition time up to this stage 
is given by
$t=\theta/F$.
This together with
Eq. (\ref{delta.vs.hf})
shows the essential property that a linear increase in the scaling range
$\Delta_0$ (given in decades) requires an exponential increase in
the duration of the experiment.
The dependence of $\Delta_0$
on the hopping energy barrier and the temperature
can be obtained from
Eq. (\ref{delta.vs.hf})
by writing $h_0$ explicitly from
(\ref{hoppingrate}) which gives

\begin{equation}
\Delta_0 = {\gamma \over 2} \left[ \log_{10} \left({ \nu \over F } \right)
- {E_0 \over k_B T} \cdot \log_{10} e  \right].
\label{delta.vs.E0T}
\end{equation}
It is easy to see that even for a system in which the energy barrier
$E_0$ vanishes, and for the extremely slow deposition rate of
$F=10^{-6}$ $ML/s$, the width of the scaling range,
assuming
$\nu=10^{12}$,
would
be
$\Delta_0 = 3$ decades.
Under these conditions, 
and taking $D=1.7$
the optimal coverage 
given by Eq. 
(\ref{maxcov})
for fractal measurement
would be
$\theta=0.126$, which would be obtained after about 35 hours
of deposition.
However, the duration of the deposition experiment in typical
submonolayer studies is usually 
limited to no more than a few hours.

The experimentally feasible scaling range is further
limited by the fact that
the diffusion properties
of physical substrates differ from the DLA model.
In particular, the assumption of an infinite separation of time scales,
namely that an isolated atom has high mobility while an atom
which has one or more nearest neighbors is completely
immobile should be weakened. 
In a real high symmetry substrate one can identify a variety of
hopping rates such as: $h_0$ for an isolated atom;
$h_{edge}$ for an atom moving along a step or island edge;
and
$h_{detach}$ for an atom detaching from a step or island edge.
We have seen that for a given substrate temperature,
the scaling range can be increased by reducing the flux $F$. 
This can be done as long as $h_0 \gg F \gg h_{edge},h_{detach}$.
However, once the duration of the experiment
(given by $t = \theta/F$) 
becomes of the order of 
$h_{edge}^{-1}$ 
or
$h_{detach}^{-1}$,
diffusion along and away from the edges becomes significant
and modifies the morphology of the islands.
These processes allow atoms to gradually diffuse into the
otherwise screened regions of the DLA-like island.
As a result, the arms becomes thicker and shorter and
the islands become more compact.
For the discussion below we will denote by 
$h_1 = \max (h_{edge},h_{detach})$
the highest hopping rate among the edge moves that
may affect the island morphology.
$h_1$ can be expressed in terms of the hopping energy 
barrier for this process, $E_1$, 
just as in
Eq.
(\ref{hoppingrate}).
The lowest deposition rate that
can be used, without having these edge processes affect the
morphology is of the order of
$F=h_1$.
Using this deposition rate 
the deposition time up to coverage $\theta$ is
$t=\theta/h_1$.
>From Eq.
(\ref{delta.vs.E0T}) 
and
$h_1 / \nu = \exp (-E_1 / k_B T)$ 
we obtain that
the maximal width 
$\Delta$
of the scaling range, in decades, 
is then given by

\begin{equation}
\Delta = {\gamma \over 2} 
\left( {E_1 - E_0 \over k_B T} \right) \log_{10} e. 
\label{width.E01}
\end{equation}
Using Eq. 
(\ref{hoppingrate})
one can eliminate the temperature and express this width in
terms of the activation energy barriers and the flux $F$
(which is chosen equal to $h_1$):

\begin{equation}
\Delta = {\gamma \over 2} \left( 1 - {E_0 \over E_1} \right)
         \log_{10} \left( {\nu \over F} \right).
\label{width.E02}
\end{equation}

To obtain the duration of the deposition experiment,
for a given $\Delta$ we extract $F$ from Eq. 
(\ref{width.E02})
and use $t=\theta/F$ where $\theta$ is given by
Eq. 
(\ref{maxcov}).
We obtain 

\begin{equation}
t = {1 \over \nu} 
10^{ K \cdot \Delta}
\label{duration}
\end{equation}
where
\begin{equation}
K= {6 E_1 \over {E_1 - E_0}} + D - 2. 
\end{equation}
This exponential dependence of the experiment duration on
$\Delta$ clearly limits the feasible scaling range
which can be obtained in these experiments.
Since $E_1 > E_0$, it is clear that $K \ge 4$. 
This lower bound is obtained for $E_0=0$ and $D=0$,
while typical values for DLA like clusters are 
$K \ge 5.7$.

Interestingly, the situation expressed by Eq.
(\ref{duration})
is somewhat reminiscent of 
that of the theory of algorithmic complexity
\cite{Garey79}.
In this theory, there is a distinction between
algorithms for which the time complexity function
depends polynomially on the input length
[typically the number of bits needed to describe the
input, i.e., $\sim \log$(input)],  
and algorithms for which the dependence is exponential.
Generally, problems for which there is a polynomial time
algorithm are considered tractable while ones for which
there are only exponential time algorithms are considered
intractable.
One can make a rough analogy between $\Delta$ and the input size, 
and the experimental duration and computation time. 
Within this analogy, the growth problem considered here, for which
the desired large value of $\Delta$ is given as input falls into
the class of intractable problems.
The understanding of
the implications of these ideas to general 
aggregation problems and other classes of fractal
systems would require further studies. 

Here we will focus on the conclusions drawn from 
Eqs. 
(\ref{width.E02})
and
(\ref{duration})
on specific experimental systems.
FCC(111) metal surfaces are the most promising experimental 
systems for studies of the growth modes considered here.
The energy barriers for Al(111) are 
$E_0=0.04$$eV$ 
and 
$E_1=0.32$$eV$
\cite{Stumpf94,Stumpf96}.
For Rh(111) 
$E_0=0.16$$eV$ 
and 
$E_1=0.54$$eV$
\cite{Ayrault74}
while for Pt(111)
$E_0=0.12$$eV$ 
and 
$E_1=0.69$$eV$
\cite{Basset78,Kellogg91}.
These numbers indicate that Al(111) can provide the
widest scaling range
for an experiment of a given duration. 
Using the equations above, for Al(111) 
we find that it is feasible
to obtain 
$\Delta=2$ decades, 
which requires $T=118K$ and $F=0.02$ $ML/s$.
However, 
$\Delta=2.5$ decades 
is already highly unfeasible since it requires a deposition
rate of about $1ML/38$ hours (at $T=94K$)!
These results seem to be consistent with the experimental 
findings reported in Section II, where for aggregation
processes no measurements are reported with 
significantly more than
two decades of scaling range.
To summarize, we have shown that the growth of DLA-like
clusters is limited by two processes: (1) the nucleation density,
and (2) edge mobility and detachment.
The resulting clusters can,
under realistic conditions,
exhibit at most 2-3 decades of scaling range.

\section{Discussion}

The MBE  systems examined here are representative in the sense 
that they exhibit spatial fractal structures
which form out of thermal equilibrium.
The need for a separation of time scales seems to be more
general for non equilibrium aggregation and growth processes,
although the details and the particular exponents may be different.
Moreover, DLA-like structures account for a significant 
fraction of the surveyed papers. The analysis presented here is
directly relevant to systems in which a finite density of
DLA-like clusters is nucleated on a substrate.
On the other hand, for growth of a single DLA-like cluster, in
problems such as electrodeposition, different considerations
are required but we believe that the issue of separation of
time scales between the fractal generating processes and the
smoothing processes determines the width of the scaling
range also there.

To explain why our arguments are 
specific to nonequilibrium systems we will use
2D percolation as an example of an equilibrium critical 
system. 
In a 2D percolation experiment, one can use a similar
apparatus as described above for MBE.
It is then assumed that diffusion is negligible and
atoms are deposited until the coverage reaches the
percolation threshold.
In such an experiment there is basically no dynamics
on the surface. 
The only constraint 
is that the deposition will be completed and
all measurements 
are performed
at a time 
scale small compared to the hopping time.
However, the hopping time can be made as long as needed by reducing the 
substrate temperature.
Under these conditions, there are no dynamical constraints
on the width of the scaling range,
which is only limited by the system size,
the precision in which the percolation threshold is approached
and the apparatus.

The discussion so far, focused on highly correlated systems
generated by dynamical processes
such as diffusion and aggregation.
However, weakly correlated systems may also exhibit fractal
behavior over a limited range of length scales.
This behavior may appear in porous media in the limit
of low volume fraction of the pores, or in surface adsorption
systems in the low coverage limit.
In this case the fractal behavior does not reflect the structure
of the basic objects (such as pores or clusters) but their distribution.
Using simple models consisting of randomly distributed spherical
or rod-like objects, we performed multiple resolution analysis
and obtained analytical expression for the box-counting function
in this case
\cite{Hamburger96a,Hamburger96b,Lidar97,Avnir97}.
It was shown that in the uncorrelated case, 
at sub-percolation coverage, 
one obtains fractal behavior over 0.5 - 2 decades.
The dimensions are found to be non-universal, and vary continuously
as a function of the coverage.
The lower cutoff in these systems is determined by the basic
object size while the upper cutoff is given by the average 
distance between them. 
It is interesting that
this independent analysis, which applies to a different
class of systems from the ones we focused on in this paper,
also gives rise to a fractal range of less than two decades.

\section{Summary}

In summary, we have performed a comprehensive survey of experimental
papers reporting fractal measurements. 
Focusing on spatial fractals, these
systems were classified according to the types of systems and processes.
It was found that for self similar fractals, the width of the scaling
range is typically limited to less than two decades with remarkably
few exceptions.
In an attempt to examine the origin of this behavior
we have focused on a class of MBE experiments in which
a finite density of DLA-like clusters nucleate and grow.
We have derived an expression of the 
duration of the deposition experiment 
which is required in order to obtain a given
width 
$\Delta$
for the scaling range. 
This expression shows that the experimental time increases
exponentially with  
$\Delta$,
given in decades.
Applying this expression to real experimental systems,
such as the MBE growth of Al on Al(111)
it is found that the feasible range is up to about two decades.
This result is in agreement with the findings of our survey
for aggregation phenomena.
Understanding the processes which determine the cutoffs in the
entire range of fractal systems,
e.g. surfaces and fronts, porous media, and other aggregation
processes requires further studies.

\acknowledgments

We would like to thank I. Furman
for helpful discussions.
This work was supported by a grant from the Wolkswagen Foundation,
administered by the Niedersachsen Science Ministry.
D.A. acknowledges support by the Minerva Foundation,
Munich.

\newpage

\begin{table}
\caption{
Experimental reports on fractals in Physical Review journals
from January 1990 to December 1996,
presented in chronological order. 
In the first column
the context of each experiment is briefly mentioned.
It is then classified, in the second column according to
the following classification: 
aggregation (A); 
porous media (P); 
surfaces and fronts (S); 
fracture (F); 
critical phenomena (C);
fracton vibrations (V);
turbulence (T);
random walk (R) and
high energy physics (H).
The next two columns provide the
fractal dimension (FD) and the width of the scaling range in which fractal
behavior was detected ($\Delta$). 
The next three columns provide the lower cutoff
($r_0$), the upper cutoff ($r_1$) and the units in which
these cutoffs are measured.
For papers in which the log-log scales are
provided in a dimensionless form or arbitrary units 
we left the units column empty.
The last two columns provide 
the Reference number and the
Figure number in that paper from which the FD, $\Delta$ and the cutoffs were
obtained.
}
\label{table.cap}
\end{table}

\newcommand{\sh}{
\scriptsize
}

\begin{center}
\begin{table}
\begin{tabular}{lcccccccc} \hline 
\sh Experiment   &\sh  Class  \ \   &\sh  FD \ \   &\sh \ \  $\Delta$ \ \  &\sh \ \ $r_0$ \ \ &\sh $r_1$ \  &\sh Units  &\sh   Ref. &\sh  Fig.   \\
             &\sh      &\sh      &\sh         &\sh             &\sh       &\sh       &\sh          &\sh        \\ \hline
\sh Aggregation of interacting colloidal gold particles &\sh A    &\sh 1.9   &\sh 1.0   &\sh 0.23   &\sh 2.3   &\sh ${\AA}^{-1}$ &\sh \cite{liu90}   &\sh 2   \\
\sh Elastic properties of colloidal gels &\sh A,P &\sh 2.0 &\sh 1.0 &\sh $0.23 \times 10^{-3}$ &\sh $2.5 \times 10^{-3}$ &\sh ${\AA}^{-1}$   &\sh \cite{shih90}  &\sh 8   \\
\sh Low frequency dynamics in superionic borate glasses &\sh V &\sh 3.3 &\sh 0.7 &\sh 1.6 &\sh 8.0 &\sh cm$^{-1}$ &\sh \cite{fontana90}  &\sh 7(a)  \\
\sh Fluctuations in granular ceramic superconductors &\sh C &\sh 2.3 &\sh 1.5 &\sh 0.027 &\sh 0.85 &\sh  &\sh \cite{ausloos90}  &\sh 2   \\  
\sh Role of local latent heat in Ge pattern formation &\sh A  &\sh 1.7 &\sh 0.7 &\sh 5.7 &\sh 28.5 &\sh  &\sh\cite{jian90}  &\sh 5   \\
\sh FD in silica aerogel - crystallized  &\sh P &\sh 2.8 &\sh 0.8 &\sh 0.8 &\sh 5.2 &\sh nm &\sh \cite{devreux90} &\sh 2   \\
\sh FD in silica aerogel - aerojel  &\sh P &\sh 2.3 &\sh 1.1 &\sh 0.13 &\sh 1.8 &\sh nm$^{-1}$ &\sh  &\sh 3   \\
\sh Vibrational dynamics in silica aerogels &\sh V &\sh 2.4 &\sh 0.9 &\sh 0.015 &\sh 0.13 &\sh ${\AA}^{-1}$ &\sh \cite{vacher90
}  &\sh 1   \\ 
\sh Conformation of graphite oxide membranes in solution &\sh S &\sh 2.4 &\sh 0.9  &\sh 2.6 &\sh 22 &\sh $\mu$m$^{-1}$ &\sh \cite{hwa91} &\sh 3  \\ 
\sh Viscous fingering in inhomogeneous porous models &\sh S &\sh 1.5 &\sh 1.3 &\sh 2.15 &\sh 40 &\sh  &\sh \cite{oxaal91} &\sh 11  \\
\sh Self-avoiding fractals: open magnetic chains in Fe-Cu &\sh R  &\sh 1.3 &\sh 1.7  &\sh 3 &\sh 148 &\sh  &\sh \cite{liu91} &\sh 2(d)  \\ 
\sh Self-avoiding fractals: closed defect loops in Ni-Mo &\sh R &\sh 1.2 &\sh 1.1 &\sh 0.023 &\sh 0.31 &\sh  &\sh   &\sh 5(b)  \\
\sh Fractal structure of cross-linked polymer resin  &\sh P &\sh 2.0 &\sh 0.7 &\sh 0.009 &\sh 0.05 &\sh ${\AA}^{-1}$  &\sh \cite{chachaty91} &\sh 1   \\
\sh Diffusion-limited-aggregation-like structures in solids  &\sh A &\sh 1.7 &\sh 1.7 &\sh 2.6 &\sh 120 &\sh  &\sh \cite{shang91} &\sh 3(a)   \\
\sh Gravity invasion percolation in 2D porous media &\sh S &\sh 1.3 &\sh 2.8 &\sh 0.05 &\sh 32 &\sh   &\sh \cite{birovljev91} &\sh 3  \\
\sh Isoscalar surfaces in turbulence &\sh S &\sh 1.7 &\sh 1.3  &\sh 5.4 &\sh 100 &\sh  &\sh \cite{constantin91}&\sh 1  \\  
\sh Viscous fingering in colloidal fluids &\sh S &\sh 1.6 &\sh 1.8  &\sh 1 &\sh 70 &\sh  &\sh\cite{lemaire91}&\sh 1(a)  \\
\sh Viscoelastic fracturing in colloidal fluids &\sh F &\sh 1.4 &\sh 1.8  &\sh 1 &\sh 70 &\sh  &\sh   &\sh 1(c)         \\ 
\sh 2D islands of Au on Ru(0001) (STM) &\sh A &\sh 1.7 &\sh 1.6  &\sh 35 &\sh 1500 &\sh $\AA$ &\sh \cite{hwang91} &\sh 4(a)  \\
\sh Hyperscaling law on polymer clusters  &\sh C &\sh 2.5 &\sh 1.0  &\sh 0.01 &\sh 0.1 &\sh ${\AA}^{-1}$ &\sh  \cite{adam91}&\sh 1  \\
\sh Structure of silica gels [light scattering(LS)] &\sh P &\sh 2.1 &\sh 1.3  &\sh $1.2\times 10^{4}$ &\sh $2.3\times 10^{5}$ &\sh cm$^{-1}$ &\sh \cite{ferri91
}&\sh 1  \\
\sh Morphology of polystyrene colloids (LS) &\sh A &\sh 2.0 &\sh 0.9 &\sh $4.3\times 10^{-4}$ &\sh $3.3\times 10^{-3}$ &\sh cm &\sh \cite{asnaghi92} &\sh 6  \\                      
\sh Morphology of polystyrene colloids &\sh A   &\sh 1.6 &\sh 1.1 &\sh $8.5\times 10^{-4}$ &\sh 0.01 &\sh cm &\sh  &\sh 7   \\
\sh Aggregation of colloidal particles at a liquid surface &\sh A &\sh 1.5 &\sh 1.6 &\sh 3.16 &\sh 112 &\sh   &\sh \cite{robinson92a} &\sh 4  \\
\sh Colloidal aggregation at the liquid-air interface &\sh A &\sh 1.6 &\sh 1.4  &\sh 1.12  &\sh 25.1 &\sh  &\sh\cite{robinson92b}&\sh 4(b)  \\
\sh Micrograph of Charpy fracture surface &\sh F &\sh 1.2 &\sh 1.9  &\sh $2.5\times 10^{-3}$ &\sh 0.22 &\sh mm &\sh  \cite{mcanulty92}&\sh 3  \\                                  
\sh Low-cycle-fatigue fracture surface &\sh F &\sh 1.4 &\sh 1.4  &\sh $2.7\times 10^{-3}$ &\sh 0.07 &\sh mm &\sh  &\sh 5  \\ 
\sh Patterns formed by laser in GeAl thin multilayer films  &\sh P &\sh  1.9 &\sh 1.5 &\sh 2 &\sh 66 &\sh  &\sh\cite{sanchez92} &\sh 2  \\
\sh Particle production in hadron-nucleus interactions &\sh H  &\sh 0.8 &\sh 1.0 &\sh 1.0 &\sh 10 &\sh  &\sh \cite{ghosh92}&\sh 3  \\
\sh Aggregation in a solution of polystyrene spheres (LS) &\sh A  &\sh 1.7 &\sh 0.7 &\sh 600 &\sh 3000 &\sh cm$^{-1}$ &\sh \cite{carpineti92} &\sh 4  \\
\sh Aggregation of self-assembled monolayer &\sh A &\sh 1.7 &\sh 1.8 &\sh 10 &\sh 600 &\sh nm &\sh\cite{schwartz92
} &\sh 4(a)  \\
\sh Infinite percolation cluster in thin films &\sh C &\sh 1.9 &\sh 1.3 &\sh 1.41 &\sh 26.6 &\sh  &\sh\cite{jensen93}&\sh 4(a)  \\
\sh Fractal dimension of fractured surface  &\sh F &\sh 1.5 &\sh 1.3 &\sh 7.5 &\sh 150 &\sh $\mu$m &\sh \cite{mu93} &\sh 1   \\
\sh Self affine growth of copper electrodeposits (STM) &\sh S  &\sh 2.5 &\sh 1.5 &\sh $10^{-4}$ &\sh $3\times 10^{-3}$ &\sh nm$^{-1}$  &\sh\cite{iwasaki93} &\sh 3  \\
\sh Growth of fractal clusters on thin solid films  &\sh A &\sh 1.7 &\sh 0.9 &\sh 7.0 &\sh 60 &\sh  &\sh \cite{zhang93}&\sh 3(a)  \\
\sh Correlations in colloidal silica aerogels  &\sh P  &\sh 1.6 &\sh 0.9 &\sh 0.3 &\sh 2.4 &\sh  &\sh \cite{hasmy93} &\sh 4(b)  \\
\sh Correlations in colloidal silica aerogels &\sh P &\sh 0.9 &\sh 0.6 &\sh 0.7 &\sh 2.8 &\sh  &\sh   &\sh 4(c)  \\
\sh Fractal electrodeposits of silver and copper films  &\sh A &\sh 1.5 &\sh 1.4 &\sh 1.0 &\sh 23 &\sh  &\sh \cite{luo93}&\sh 2(c)  \\
\sh Multifractal analysis of nucleus-nucleus interactions &\sh H  &\sh 1.0 &\sh 1.0 &\sh 1.0 &\sh 10 &\sh  &\sh \cite{sengupta93}&\sh 2  \\
\sh Period-doubling scenarios in Taylor-Couette flow &\sh T &\sh 2.4 &\sh 1.4 &\sh 2.0 &\sh 45 &\sh  &\sh \cite{buzug93}&\sh 9(a)  \\
\sh 2D aggregation of polystyrene latex particles (optical)&\sh A &\sh 1.5 &\sh 1.8 &\sh 0.56 &\sh 31.6 &\sh  &\sh \cite{stankiewicz93} &\sh 2  \\
\sh Nucleation-limited aggregation in aqueous-solution films (STM) &\sh A &\sh 1.8 &\sh 1.6 &\sh  5.0 &\sh 220 &\sh  &\sh\cite{ming93} &\sh 1(b)   \\
\sh Fractal electrodeposits grown under damped free convection &\sh A &\sh 2.5 &\sh 1.2 &\sh 0.06 &\sh 0.87 &\sh cm &\sh\cite{carro93}&\sh 3(a)   \\
\sh Colloidal aggregation induced by alternating electric fields &\sh A &\sh 1.5 &\sh 1.4 &\sh 1.8 &\sh 42 &\sh $\mu$m &\sh \cite{wei93} &\sh 2(b)  \\
\sh Fractal electrodes and interfaces &\sh S &\sh 2.4 &\sh 3.8 &\sh 10 &\sh $6\times 10^{4}$ &\sh Hz &\sh\cite{sapoval93}&\sh 13  \\
\sh Fractal distribution of earthquake hypocenters &\sh F  &\sh 1.8 &\sh 1.4 &\sh 5.0 &\sh 120 &\sh km &\sh \cite{sahimi93}&\sh 3  \\
\sh Pore space correlations in capillary condensation (LS) &\sh P  &\sh 2.6 &\sh 1.4 &\sh 0.1 &\sh 2.5 &\sh $\mu$m$^{-1}$ &\sh \cite{page93
} &\sh 3  \\
\sh Water desorption and adsorption in porous materials &\sh P &\sh 1.7 &\sh 0.8 &\sh 0.02 &\sh 0.14 &\sh ${\AA}^{-1}$ &\sh \cite{li94} &\sh 3  \\
\sh Spin-lattice relaxation by paramagnetic dopants in $Li{\footnotesize _2}Si{\footnotesize _2}O{\footnotesize _5}$ &\sh C  &\sh  3.0     &\sh  1.3    &\sh 0.5 &\sh 10 &\sh s &\sh   \cite{sen94}   &\sh  5   \\
\sh Spin-lattice relaxation by paramagnetic dopants in $Na{\footnotesize _2}Si{\footnotesize _2}O{\footnotesize _5}$ &\sh C &\sh 2.1 &\sh 1.3 &\sh 0.5 &\sh 10 &\sh s  &\sh   &\sh 5  \\
\sh Interface thickness in block copolymers &\sh S  &\sh 2.5 &\sh 0.9 &\sh 0.03 &\sh 0.25 &\sh ${\AA}^{-1}$ &\sh \cite{xie94} &\sh 8(a)  \\ 
\sh Long range correlations in Silica aerogels &\sh A,P &\sh 1.7 &\sh 1.1 &\sh 0.015 &\sh 0.2 &\sh  ${\AA}^{-1}$ &\sh  \cite{hasmy94}&\sh 10  \\
\sh Low-frequency vibrational states in $As{\small _2}S{\small _3}$ glass &\sh V &\sh 2.4 &\sh 0.4 &\sh 14 &\sh 32 &\sh cm$^{-1}$ &\sh \cite{ohsaka94} &\sh 3  \\
\sh Heavily irradiated pure and doped NaCl crystals (Raman)&\sh P  &\sh 2.5 &\sh 1.2 &\sh 6 &\sh 100 &\sh  cm$^{-1}$ &\sh \cite{groote94}&\sh 1  \\
\sh Multihadron production in high energy interactions &\sh H &\sh 0.9 &\sh 1.0 &\sh 1.0 &\sh 10 &\sh  &\sh\cite{ghosh94}&\sh 2  \\
\sh Pseudorapidity distribution for particles produced in pp collisions. &\sh H &\sh 1.0 &\sh 1.3 &\sh 0.5 &\sh 10 &\sh  &\sh \cite{shaoshun94}&\sh 2  \\
\sh Multifractal moments in 800 GeV proton-nucleus interactions &\sh H &\sh 0.7 &\sh 1.7 &\sh 0.2 &\sh 9 &\sh  &\sh \cite{shivpuri94
}&\sh 1(a)  \\
\sh Electrodeposition of a gold oxide layer on a gold cathode (STM)&\sh S  &\sh 2.2 &\sh 1.5 &\sh 40 &\sh 1258 &\sh $\AA$ &\sh \cite{vazquez94}&\sh 4(a)  \\
\sh Aggregation of 2D polystyrene particles (in-situ microscopy) &\sh A &\sh 1.8 &\sh 1.3 &\sh 10 &\sh 220 &\sh $\mu$m &\sh \cite{wei94}&\sh 3(d)  \\
\sh Fractal scaling behavior of vapor-deposited silver films &\sh S &\sh 2.4 &\sh 0.6 &\sh 40 &\sh 150 &\sh  &\sh \cite{penella94}&\sh 3  \\
\sh Tracer dispersion fronts in porous media (computer imaging) &\sh S  &\sh 1.4 &\sh 2.5 &\sh 0.1 &\sh 32 &\sh  &\sh \cite{birovljev94} &\sh 5  \\
\sh Teritary structure of proteins &\sh R &\sh 1.6 &\sh 1.3 &\sh 50 &\sh 1000 &\sh  &\sh \cite{xiao94}&\sh 1  \\
\sh Dense colloid silica suspensions in a $H{\footnotesize _2}O-D{\footnotesize _2}O$ medium &\sh P &\sh 1.6 &\sh 0.4 &\sh 0.9 &\sh 2.5 &\sh  &\sh \cite{muzny94} &\sh 2  \\
\sh 2D aluminum corrosion fronts &\sh S  &\sh 1.2 &\sh 3.7 &\sh 2.0 &\sh $10^{4}$ &\sh $\mu$m &\sh \cite{holten94} &\sh 4  \\
\sh Aggregation of polystyrene latices (LS) &\sh A &\sh 1.7 &\sh 0.8 &\sh 100 &\sh 600 &\sh nm &\sh \cite{zhu94} &\sh 4(a)  \\
\sh Aggregation of polystyrene latices (LS) &\sh A  &\sh 2.7 &\sh 0.5 &\sh 200 &\sh 630 &\sh nm &\sh   &\sh 4(c)  \\
\sh Diffusion of aggregates in carbonaceous flame soot aerosol (LS)&\sh A,P  &\sh 2.2 &\sh 0.4 &\sh 2.0 &\sh 5.0 &\sh  &\sh \cite{cai94} &\sh 2  \\
\sh Spinodal decomposition in hydrogen-bonded polymer &\sh A &\sh 2.4 &\sh 0.4 &\sh $5.6\times 10^{-3}$ &\sh $15\times 10^{-3}$ &\sh  &\sh \cite{hobbie94} &\sh 3(a)  \\
\sh Broadband edge density fluctuations in compact helical system &\sh T &\sh 6.0 &\sh 2.0 &\sh 100 &\sh $10^{4}$ &\sh  &\sh \cite{komori94} &\sh 3(a)  \\
\sh Graphitic oxide sheets suspended in aqueous solution          &\sh F &\sh 2.1 &\sh 1.1 &\sh 2.0 &\sh 25 &\sh $\mu$m$^{-1}$ &\sh \cite{spector94} &\sh 2  \\
\sh Structural analysis of electroless deposits &\sh A  &\sh 1.6 &\sh 1.3 &\sh 0.05 &\sh 1.0 &\sh  &\sh \cite{kuhn94
}&\sh 5(b)  \\
\sh Boson peak in the raman spectra of amorphous GaAs &\sh V &\sh 2.5 &\sh 0.6 &\sh 300 &\sh 1200 &\sh cm$^{-1}$ &\sh \cite{ivanda95}&\sh 5 \\
\sh Fractal structure of porous solides characterized by adsorption &\sh P  &\sh 2.6 &\sh 0.4 &\sh 5.6 &\sh 12.6 &\sh  &\sh \cite{sze95}&\sh 1(b) \\
\sh Cold deposited silver flms determined by low temperature STM &\sh S &\sh 2.5 &\sh 1.8 &\sh0.03 &\sh 2.0 &\sh nm$^{-1}$ &\sh \cite{douketis95}&\sh6(a) \\
\sh Porous glass characterized by adsorbed dibromomethane  &\sh P  &\sh 2.3 &\sh 0.7 &\sh 0.03 &\sh 0.15 &\sh ${\AA}^{-1}$ &\sh \cite{mitropoulos95}&\sh 3 \\
\sh Multifractality of medium energy particles in p-AgBr interactions &\sh H &\sh 0.7 &\sh 0.6 &\sh 1.22 &\sh 4.95 &\sh  &\sh \cite{shivpuri95}&\sh 2(a) \\ 
\sh Multifractality in proton-nucleus interaction &\sh H &\sh 0.9 &\sh 1.3 &\sh 2 &\sh 44 &\sh  &\sh \cite{dheer95} &\sh 3  \\
\sh Multiplicity distributions from central collisions ${}^{16}$O$+$Cu &\sh H  &\sh 1.0 &\sh 1.0 &\sh 1.0 &\sh 10 &\sh  &\sh \cite{abbott95} &\sh 8(a) \\
\sh Fractal analysis of the multiparticle production process &\sh H &\sh 0.8 &\sh 1.0 &\sh 4.0 &\sh 40 &\sh  &\sh \cite{ghosh95} &\sh 7 \\
\sh Double layer relaxation at rough electrodes &\sh A &\sh 2.5 &\sh 0.5 &\sh 0.3 &\sh 1.0 &\sh $\mu$A &\sh \cite{larsen95}&\sh 2 \\
\sh Long range correlations in DNA sequences from wavelet analysis &\sh R &\sh 1.0 &\sh 2.4 &\sh 16 &\sh 4100 &\sh  &\sh  \cite{arneodo95} &\sh 2 \\
\sh Percolation in a 3D disordered conductor insulator composite  &\sh C &\sh 1.9 &\sh 0.8 &\sh 0.1 &\sh 0.6 &\sh $\mu$m &\sh \cite{viswanathan95
}&\sh 3 \\ 
\sh Percolation in a 3D disordered conductor insulator composite  &\sh C &\sh 2.6 &\sh 0.5 &\sh 0.6 &\sh 2.0 &\sh $\mu$m &\sh &\sh 3 \\
\sh Oxide aggregation on liquid-gallium surface &\sh A &\sh 1.5 &\sh 2.1 &\sh 0.45 &\sh 55 &\sh $\mu$m &\sh \cite{wang96} &\sh 4 \\
\sh Dense branching morphology in Bi/Al/Mn/SiO films  &\sh S  &\sh 1.6 &\sh 2.0 &\sh $9\times 10^{-3}$ &\sh 1.0 &\sh  &\sh\cite{shang96}&\sh 11(a) \\
\sh Evolution of source rocks during hydrocarbon generation &\sh P &\sh 2.5 &\sh 1.6 &\sh $5\times 10^{-3}$ &\sh 0.2 &\sh ${\AA}^{-1}$ &\sh \cite{radlinski96}&\sh  4(b) \\ 
\sh Fractal dimension of Li insertion electrodes &\sh S &\sh 2.3 &\sh 2.0 &\sh 5.0 &\sh 500 &\sh mV/s &\sh \cite{mattsson96a}&\sh 2 \\
\sh cyclic I-V studies of In oxide films &\sh S &\sh 1.8 &\sh 2.3 &\sh 1.0 &\sh 200 &\sh mV/s &\sh  \cite{mattsson96b} &\sh 2 \\
\sh Sn oxyfluoride               &\sh S &\sh 1.9 &\sh 1.2 &\sh 1.5 &\sh 23 &\sh $\mu$m &\sh   &\sh 4 \\
\sh Intermittency in ${}^{197}$Au fragmentation &\sh H &\sh 1.0 &\sh 1.3 &\sh 2.0 &\sh 40 &\sh   &\sh \cite{cherry96}&\sh 4(a) \\
\sh Evaporatively controlled growth of salt trees &\sh A &\sh 2.3 &\sh 0.8 &\sh 0.25 &\sh 1.6 &\sh cm &\sh \cite{du96}&\sh 4(a) \\
\sh Fractal growth during annealing of aluminum on silica &\sh A &\sh 1.7 &\sh 2.3 &\sh 1.0 &\sh 200 &\sh  &\sh \cite{balazs96} &\sh 5 \\
\sh Flow of water pumped through pore space (NMR)   &\sh P,C &\sh 1.8 &\sh 0.5 &\sh 1.0 &\sh 3.5 &\sh  &\sh \cite{muller96} &\sh 7(a) \\
\sh Formation of side branches of xenon dendrites    &\sh S &\sh 1.4 &\sh 2.4 &\sh 0.015 &\sh 4.0 &\sh mm &\sh \cite{bisang96} &\sh 11 \\
\sh Aggregation of porphyrins in aqueous solutions &\sh A &\sh 2.5 &\sh 1.7 &\sh 0.65 &\sh 30 &\sh $\mu$m$^{-1}$ &\sh \cite{mallamace96}&\sh 1 \\
\sh Structure and Pertinent length scale of discotic clay gel &\sh P &\sh 1.8 &\sh 0.9 &\sh $2\times 10^{-5}$ &\sh $1.5\times 10^{-4}$ &\sh ${\AA}^{-1}$ &\sh   \cite{pignon96} &\sh 1(a) \\

\end{tabular}
\end{table}
\end{center}

\newpage

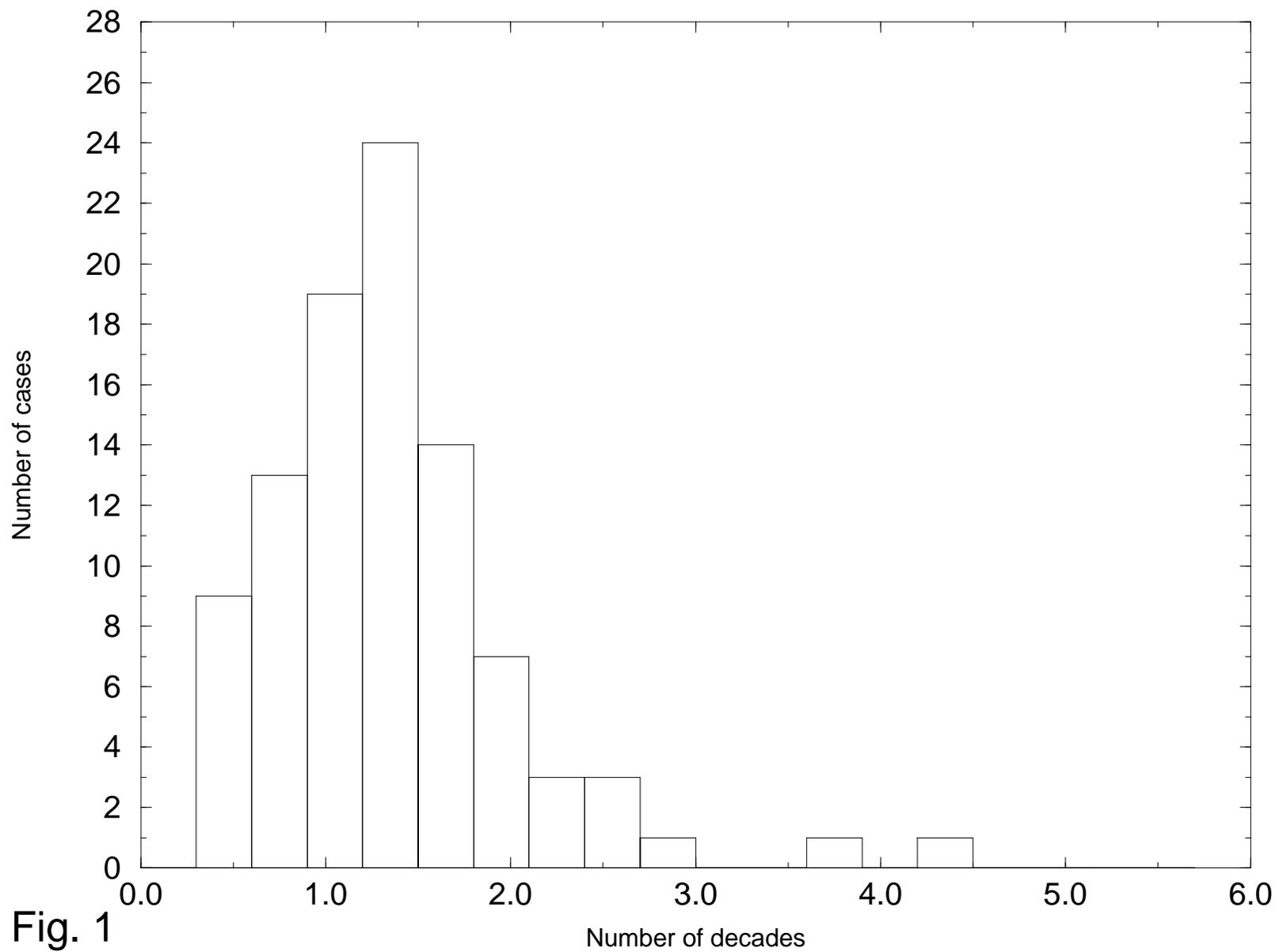
\begin{figure}
\caption{
Distribution of the widths of the scaling range 
for fractal measurements reported in
{\it Physical Review} journals between 1990 and 1996.
The horizontal axis shows
the width of the linear range in the log-log plots
(measured in decades) 
over which the FD was determined and the vertical axis shows
the number of measurements in which a given width was obtained. 
Note that most fractal measurements appear to be based on data that extends
between 0.5 - 2 decades.
The bin-width is 0.3 decade.
}
\label{width1.hist}
\end{figure}

\begin{figure}
\caption{
The distributions of the widths of the scaling range
for particular classes of spatial fractals:
(a) aggregation; (b) porous media and (c) surfaces and fronts.
}
\label{width2.hist}
\end{figure}

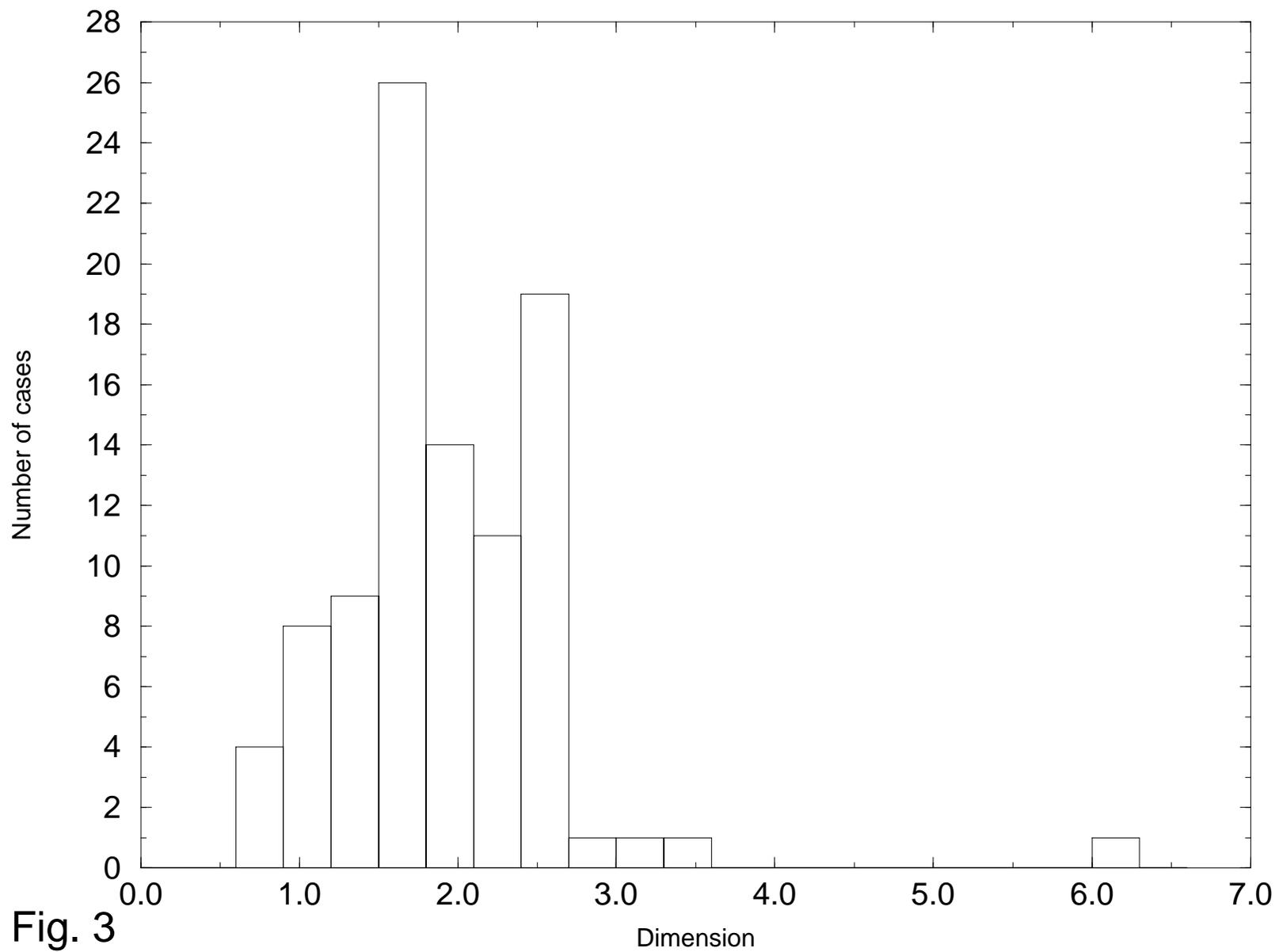
\begin{figure}
\caption{ 
Distribution of experimentally measured fractal dimensions. 
A broad distribution is observed with
peaks around $D=1.7$ and $D=2.5$. 
The bin-width is 0.3.
}
\label{dim1.hist}
\end{figure}

\begin{figure}
\caption{ 
The distributions of fractal dimensions for particular 
classes of spatial fractals:
(a) aggregation; (b) porous media and (c) surfaces and fronts.
}
\label{dim2.hist}
\end{figure}


\begin{thebibliography}{10}


\bibitem{Mandelbrot82}
{B. B. Mandelbrot}, {\em {The Fractal Geometry of Nature}} ({Freeman}, {San
  Francisco}, 1982).


\bibitem{Falconer90}
K. Falconer,
{\it Fractal Geometry: Mathematical Foundations and Applications}
(Wiley, Chichester, 1990).


\bibitem{Stanley86}
{\em On Growth and Form}, No.~100 in {\em NATO ASI Ser. E}, edited by {H. E.
  Stanley, N. Ostrowsky} (Martinus Nijhoff, Dordrecht, 1986).

\bibitem{Vicsek89}
T. Vicsek, {\it Fractal Growth Phenomena},
(World Scientific, Singapore, 1989).

\bibitem{Feder90}
{\em {Fractals in Physics, Essays in Honour of B.B. Mandelbrot}}, edited by {J.
  Feder and A. Aharony} ({North Holland}, {Amsterdam}, 1990).

\bibitem{Takayasu90}
{H. Takayasu}, {\em {Fractals in the Physical Sciences}} ({J. Wiley \& Sons},
  {Chichester}, 1990).


\bibitem{Avnir92}
{\em {The Fractal Approach to Heterogeneous Chemistry: Surfaces, Colloids,
  Polymers}}, edited by {D. Avnir} ({John Wiley \& Sons Ltd.}, Chichester,
  1992).

\bibitem{Bunde94}
{\em Fractals in Science}, edited by {A. Bunde, S. Havlin} (Springer, Berlin,
  1994).

\bibitem{Barabasi95}
{A.-L. Barab\'{a}si and H.E. Stanley}, {\em {Fractal Concepts in Surface
  Growth}} ({Cambridge University Press}, {Cambridge}, 1995).

\bibitem{Stanley71}
H.E. Stanley,
{\it Introduction to Phase Transitions and Critical Phenomena}
(Oxford University Press, Oxford, 1971).


\bibitem{Binney92}
J.J. Binney, N.J. Dowrick, A.J. Fisher and M.E.J. Newman,
{\it The Theory of Critical Phenomena: An Introduction
to the Renormalization Group}
(Clarendon Press, Oxford, 1992).

\bibitem{Kapitulnik84a}
A. Kapitulnik and G. Deutscher,
{\it J. Stat. Phys.} {\bf 36}, 815 (1984).

\bibitem{Kapitulnik84b}
A. Kapitulnik, Y. Gefen and A. Aharony,
{\it J. Stat. Phys.} {\bf 36}, 807 (1984).

\bibitem{Isichenko92}
M.B. Isichenko,
{\it Rev. Mod. Phys.} {\bf 64}, 961 (1992).

\bibitem{Stauffer79}
D. Stauffer,
{\it Physics Reports} {\bf 54}, 1 (1979).

\bibitem{Stauffer85}
D. Stauffer,
Introduction to Percolation Theory 
(Taylor and Francis, London, 1985)

\bibitem{Eckmann85}
J.-P. Eckmann and D. Ruelle,
{\it Rev. Mod. Phys.} {\bf 57}, 617 (1985).

\bibitem{Stavans85}
J. Stavans, F. Heslot and A. Libchaber,
{\it Phys. Rev. Lett.} {\bf 55}, 596 (1985).

\bibitem{Jensen85}
M.H. Jensen, L.P. Kadanoff, A. Libchaber, I. Procaccia
and J. Stavans,
{\it Phys. Rev. Lett.} {\bf 55}, 2798 (1985).

\bibitem{Grassberger84}
P. Grassberger and I. Procaccia,
{\it Physica} {\bf D13}, 34 (1984).

\bibitem{Eckmann92}
J.-P. Eckmann and D. Ruelle,
{\it Physica} {\bf D56}, 185 (1992).

\bibitem{Pfeifer92}
P. Pfeifer and M. Obert, in Ref. 
\cite{Avnir92}, p. 11.

\bibitem{Witten81}
T. A. Witten and L. M. Sander,
{\it Phys. Rev. Lett.} {\bf 47}, 1400 (1981).

\bibitem{Witten83}
T. A. Witten and L. M. Sander,
{\it Phys. Rev.} {\bf B27}, 5686 (1983).

\bibitem{Meakin88}
P. Meakin, in {\it Phase Transitions and Critical Phenomena},
Ed. C. Domb and J.L. Lebowitz, {\bf 12} 334 (1988).


\bibitem{Mandelbrot92}
B.B. Mandelbrot,
{\it Physica} {\bf A191}, 95 (1992).

\bibitem{Eckmann90}
J.-P. Eckmann, P. Meakin, I. Procaccia and R. Zeitak,
{\it Phys. Rev. Lett.} {\bf 65}, 52 (1990);

\bibitem{Arneodo92}
A. Arneodo, F. Argoul, J.F. Muzy and M. Tabard,
{\it Physica} {\bf A188}, 217 (1992).

%3D dla - dimension by numerical simulation.
\bibitem{Meakin85}
P. Meakin,
{\it Phys. Rev.} {\bf A27}, 604 (1985).

\bibitem{Brady84}
R.M. Brady and R.C. Ball,
{\it Nature} {\bf 309}, 225 (1984).


\bibitem{Hwang91}
R. Q. Hwang, J. Schr\"{o}der, C. G\"{u}nter, and
R. J. Behm, {\it Phys. Rev. Lett.} {\bf 67},
3279 (1991).

\bibitem{Struzik97}
Z.R. Struzik, E.H. Dooijes and F.C.A. Groen,
{\it Fractal Frontiers}, 
Edited by M.M. Novak and T.G. Dewey
(World Scientific, Singapore, 1997), p. 189.


\bibitem{melvyl}
The search was done using the command
``find kw fractal or fractals and date 199n and jo
physical review and pt experimental'' for $n=0,\dots6$.
The numbers of papers obtained were
$23$, $16$, $19$, $28$, $30$, $23$ and $26$ for $1990,\dots,1996$,
respectively, a total of 165 papers. 


% ******** Below is the list of references for the Table ******

\bibitem{liu90}
		J. Liu, W. Y. Shih, M. Sarikaya and I. A. Aksay,
		{\it Phys. Rev. } {\bf A41}, 3206 (1990).
\bibitem{shih90}
		W. -H. Shih, W. Y. Shih, S. -I. Kim, J. Liu and I. A. Aksay,
		{\it Phys. Rev. } {\bf A42}, 4772 (1990).

\bibitem{fontana90}
		A. Fontana, F. Rocca, 
                M. P. Fontana, B. Rosi and A. J. Dianoux,
		{\it Phys. Rev.} {\bf B41}, 3778 (1990).

\bibitem{ausloos90}
		M. Ausloos and P. Clippe,
		{\it Phys. Rev. } {\bf B41}, 9506 (1990).

\bibitem{jian90}
		H. Jian-guo and W. Zi-qin,
		{\it Phys. Rev. } {\bf B42}, 3271 (1990).

\bibitem{devreux90}
		F. Devreux, J. P. Boilot, F. Chaput, and
		B. Sapoval, {\it Phys. Rev. Lett.} {\bf 65},
		614 (1990).

\bibitem{vacher90
}
		R. Vacher, E. Courtens, G. Coddens, A. Heidemann,
		Y. Tsujimi, J. Pelous and M. Foret, 
                {\it Phys. Rev. Lett.} {\bf 65}, 1008 (1990).

\bibitem{hwa91}
		T. Hwa, E. Kokufuta and T. Tanaka,
		{\it Phys. Rev. } {\bf A44}, R2235 (1991).

\bibitem{oxaal91}
		U. Oxaal,
		{\it Phys. Rev. } {\bf A44}, 5038 (1991).

\bibitem{liu91}
	        B. X. Liu, C. H. Shang and H. D. Li,
		{\it Phys. Rev. } {\bf B44}, 4365 (1991).

\bibitem{chachaty91}
		C. Chachaty, J. -P. Korb, J. R. C. van der Maarel, W. Bras and P. Quinn.
		{\it Phys. Rev. } {\bf B44}, 4778 (1991).

\bibitem{shang91}
	        C. H. Shang, B. X. Liu, J. G. Sun and H. D. Li,
		{\it Phys. Rev. } {\bf B44}, 5035 (1991).

\bibitem{birovljev91}
		A. Birovljev, L. Furuberg, J. Feder, T. J{\o}ssang,
                K. M{\aa}l{\o}y and A. Aharony, 
                {\it Phys. Rev. Lett.} {\bf 67}, 584 (1991).).

\bibitem{constantin91}
		P. Constantin, I. Procaccia and K. R. Sreenivasan,
                {\it Phys. Rev. Lett.} {\bf 67}, 1739 (1991).

\bibitem{lemaire91}
		E. Lemaire, P. Levitz, G. Daccord and H. Van Damme,
                {\it Phys. Rev. Lett.} {\bf 67}, 2009 (1991).

\bibitem{hwang91}
		R. Q. Hwang, J. Schr\"{o}der, C. G\"{u}nther, and
		R. J. Behm, {\it Phys. Rev. Lett.} {\bf 67},
		3279 (1991).

\bibitem{adam91}
		M. Adam, D. Lairez, F. Bou{\'{e}}, J. P. Busnel,
                D. Durand and T. Nicolai, {\it Phys. Rev. Lett.} {\bf 67},
                3456 (1991).

\bibitem{ferri91
}
		F. Ferri, B. J. Frisken and D. S. Cannell,
		{\it Phys. Rev. Lett.} {\bf 67}, 3626 (1991).

\bibitem{asnaghi92}
		D. Asnaghi, M. Carpineti, M. Giglio and M. Sozzi,
		{\it Phys. Rev. } {\bf A45}, 1018 (1992).

\bibitem{robinson92a}
		D. J. Robinson and J. C. Earnshaw,
		{\it Phys. Rev. } {\bf A46}, 2045 (1992).

\bibitem{robinson92b}
		D. J. Robinson and J. C. Earnshaw,
		{\it Phys. Rev. } {\bf A46}, 2065 (1992).

\bibitem{mcanulty92}
		P. McAnulty, L. V. Meisel and P. J. Cote,
		{\it Phys. Rev. } {\bf A46}, 3523 (1992).

\bibitem{sanchez92}
	        A. S{\'{a}}nchez, R. Serna, F. Catalina and C. N. Afonso,
		{\it Phys. Rev. } {\bf B46}, 487 (1992).

\bibitem{ghosh92}
		D. Ghosh, P. Ghosh, A. Deb, D. Halder, S. Das, A. Hossain,
		A. Dey and J Roy, {\it Phys. Rev. } {\bf D46}, 3712 (1992).

\bibitem{carpineti92}
		M. Carpineti and M. Giglio,
                {\it Phys. Rev. Lett.} {\bf 68}, 3327 (1992).

\bibitem{schwartz92
}
		D. K. Schwartz, S. Steinberg, J. Israelachvili, and J. A. N. Zasadzinski,
                {\it Phys. Rev. Lett.} {\bf 69}, 3354 (1992).

\bibitem{jensen93}
		P. Jensen, P. Melinon, M. Treilleux, J. X. Hu,
		J. Dumas, A. Hoareau and B. Cabaud,{\it Phys. Rev. } {\bf B47},
                5008 (1993).

\bibitem{mu93}
		Z. Q. Mu, C. W. Lung, Y. Kang and Q. Y. Long,
		{\it Phys. Rev. } {\bf B48}, 7679 (1993).

\bibitem{iwasaki93}
	        H. Iwasaki and T. Yoshinobu,
		{\it Phys. Rev. } {\bf B48}, 8282 (1993).

\bibitem{zhang93}
		J. Zhang, D. Liu and K. Colbow,
		{\it Phys. Rev. } {\bf B48}, 9130 (1993).

\bibitem{hasmy93}
		A. Hasmy, M. Foret, J. Pelous and R. Jullien,
		{\it Phys. Rev. } {\bf B48}, 9345 (1993).

\bibitem{luo93}
		G. P. Luo, Z. M. Ai, J. J. Hawkes, Z. H. Lu and Y. Wei,
		{\it Phys. Rev. } {\bf B48}, 15337 (1993).

\bibitem{sengupta93}
		K. Sengupta, M. L. Cherry, W. V. Jones, J. P. Wefel, A. Dabrowska,
                R. Holy{\'{n}}ski, A. Jurak, A. Olszewski, M. Szarska, A. Trzupek, 
                B. Wilczy{\'{n}}ska,
                H. Wilczy{\'{n}}ski, W. Wolter, B. Wosiek, K. Wo{\'{z}}niak,
                P. S. Freier and C. J. Waddington, {\it Phys. Rev. } {\bf D48}, 
                3174 (1993).

\bibitem{buzug93}
		T. Buzug, J. von Stamm and G. Pfister,
		{\it Phys. Rev. } {\bf E47}, 1054 (1993).

\bibitem{stankiewicz93}
		J. Stankiewicz, M. A. C. Vilchez and R. H. Alvarez, 
		{\it Phys. Rev. } {\bf E47}, 2663 (1993).

\bibitem{ming93}
		N. Ming, M. Wang and R.-W. Peng
		{\it Phys. Rev. } {\bf E48}, 621 (1993).

\bibitem{carro93}
		P. Carro, S. L. Marchiano, A. H. Creus, S. Gonz{\'{a}}lez,
	        R. C. Salvarezza and A. J. Arvia, {\it Phys. Rev. } {\bf E48},
                R2374 (1993).

\bibitem{wei93}
		Q. Wei, X. Liu, C. Zhou and N. Ming,
		{\it Phys. Rev. } {\bf E48}, 2786 (1993).

\bibitem{sapoval93}
		B. Sapoval, R. Gutfraind, P. Meakin, M. Keddam, H. Takenouti,
		{\it Phys. Rev. } {\bf E48}, 3333 (1993).

\bibitem{sahimi93}
		M. Sahimi, M. C. Robertson and C. G. Sammis,
                {\it Phys. Rev. Lett.} {\bf 70}, 2186 (1993).

\bibitem{page93
}
		J. H. Page, J. Liu, B. Abeles, H. W. Deckman and D. A. Weitz,
                {\it Phys. Rev. Lett.} {\bf 71}, 1216 (1993).

\bibitem{li94}
		J. C. Li, D. K. Ross, L. D. Howe, K. L. Stefanopoulos, J. P. A. Fairclough, 
	        R. Heenan, and K. Ibel, {\it Phys. Rev. } {\bf B49}, 5911 (1994).

\bibitem{sen94}
		S. Sen and J. F. Stebbins,
		{\it Phys. Rev. } {\bf B50}, 822 (1994).

\bibitem{xie94}
		R. Xie, B. Yang, and B. Jiang, 
	        {\it Phys. Rev. } {\bf B50}, 3636 (1994).

\bibitem{hasmy94}
		A. Hasmy, E. Anglaret, M. Foret, J. Pelous, and R. Jullien,
		{\it Phys. Rev. } {\bf B50}, 6006 (1994).

\bibitem{ohsaka94}
		T. Ohsaka and T. Ihara,  
		{\it Phys. Rev. } {\bf B50}, 9569 (1994).

\bibitem{groote94}
		J. C. Groote, J. R. W. Weerkamp, J. Seinen, and H. W. den Hartog,
		{\it Phys. Rev. } {\bf B50}, 9798 (1994).

\bibitem{ghosh94}
		D. Ghosh, A. Deb, M. Lahiri, A. Dey, S. A. Hossain, S. Das,
		S. Sen and S. Halder,{\it Phys. Rev. } {\bf D49}, 3113 (1994)

\bibitem{shaoshun94}
		W. Shaoshun, Z. Jie, Y. Yunxiu, X. Chenguo and Z. Yu,  
		{\it Phys. Rev. } {\bf D49}, 5785 (1994).

\bibitem{shivpuri94
}
		R. K. Shivpuri and V. Anand,  
		{\it Phys. Rev. } {\bf D50}, 287 (1994).

\bibitem{vazquez94}
		L. V{\'{a}}zquez,  R. C. Salvarezza, P. Oc{\'{o}}n, 
                P. Herrasti,  
		J. M. Vara and A. J. Arvia, {\it Phys. Rev. } {\bf E49},
                1507 (1994).

\bibitem{wei94}
		Q. Wei, M. Han, C. Zhou and N. Ming,
		{\it Phys. Rev. } {\bf E49}, 4167 (1994).

\bibitem{penella94}
		V. Panella and J. Krim,  
		{\it Phys. Rev. } {\bf E49}, 4179 (1994).

\bibitem{birovljev94}
		A. Birovljev, K. J. M{\aa}l{\o}y, J. Feder and T. J{\o}ssang,    
		{\it Phys. Rev. } {\bf E49}, 5431 (1994).

\bibitem{xiao94}
		Y. Xiao,  
		{\it Phys. Rev. } {\bf E49}, 5903 (1994).

\bibitem{muzny94}
		C. D. Muzny, G. C. Straty and H. J. M. Hanley,  
		{\it Phys. Rev. } {\bf E50}, R675 (1994).

\bibitem{holten94}
		T. Holten, T. J{\o}ssang, P. Meakin ang J. Feder,  
		{\it Phys. Rev. } {\bf E50}, 754 (1994).

\bibitem{zhu94}
		P. W. Zhu and D. H. Napper,  
		{\it Phys. Rev. } {\bf E50}, 1360 (1994).

\bibitem{cai94}
		J. Cai and C. M. Sorensen,
		{\it Phys. Rev. } {\bf E50}, 3397 (1994).

\bibitem{hobbie94}
	        E. K. Hobbie, B. J. Bauer and C. C. Han,
                {\it Phys. Rev. Lett.} {\bf 72}, 1830 (1994).

\bibitem{komori94}
	        A. Komori, T. Baba, T. Morisaki, M. Kono, H. Iguchi, K. Nishimura, H. Yamada, 
                S. Okamura, and K. Matsuoka, {\it Phys. Rev. Lett.} {\bf 73}, 660 (1994).

\bibitem{spector94}
	        M. S. Spector, E. Naranjo, S. Chiruvolu, and J. A. Zasadzinski, 
                {\it Phys. Rev. Lett.} {\bf 73}, 2867 (1994).

\bibitem{kuhn94
}
	        A. Kuhn, F. Argoul, J. F. Muzy, and A. Arneodo, 
                {\it Phys. Rev. Lett.} {\bf 73}, 2998 (1994).

\bibitem{ivanda95}
		M. Ivanda, I. Hartmann, W. Kiefer,
	        {\it Phys. Rev. } {\bf B51}, 1567 (1995).

\bibitem{sze95}
		S. J. Sze and T. Y. Lee,
	        {\it Phys. Rev. } {\bf B51}, 8709 (1995).

\bibitem{douketis95}
		C. Douketis, Z. Wang, T. L. Haslett, and M. Moskovits,
	        {\it Phys. Rev. } {\bf B51}, 11022 (1995).

\bibitem{mitropoulos95}
		A. C. Mitropoulos, J. M. Haynes, and R. M. Richardson, 
	        {\it Phys. Rev. } {\bf B52}, 10035 (1995).

\bibitem{shivpuri95}
		R. K. Shivpuri, G. Das, and S. Dheer,
	        {\it Phys. Rev. } {\bf C51}, 1367 (1995).

\bibitem{dheer95}
		S. Dheer, G. Das, R. K. Shivpuri, and S. K. Soni,
	        {\it Phys. Rev. } {\bf C52}, 1572 (1995).

\bibitem{abbott95}
		T. Abbott et al.,  
	        {\it Phys. Rev. } {\bf C52}, 2663 (1995).

\bibitem{ghosh95}
		D. Ghosh, A. Deb, and M. Lahiri,
	        {\it Phys. Rev. } {\bf D51}, 3298 (1995).

\bibitem{larsen95}
		A. E. Larsen, D. G. Grier, and T. C. Halsey,
	        {\it Phys. Rev. } {\bf E52}, R2161 (1995).

\bibitem{arneodo95}
	        A. Arneodo, E. Bacry, P. V. Graves, and J. F. Muzy,
	        {\it Phys. Rev. Lett} {\bf 74}, 3293 (1995).

\bibitem{viswanathan95
}
		R. Viswanathan, and M. B. Heaney, 
	        {\it Phys. Rev. Lett.} {\bf 75}, 4433 (1995).


\bibitem{wang96} 
                Y.L. Wang and S.J. Lin,
	        {\it Phys. Rev.} {\bf B53}, 6152 (1996).

\bibitem{shang96}
		C. H. Shang,
	        {\it Phys. Rev. } {\bf B53}, 13759 (1996).

\bibitem{radlinski96}
		A. P. Radlinski and C. J. Boreham,
	        {\it Phys. Rev. } {\bf B53}, 14152 (1996).




\bibitem{mattsson96a}
		M. S. Mattsson, G. A. Niklasson, and C. G. Granqvist
	        {\it Phys. Rev. } {\bf B54}, 2968 (1996).

\bibitem{mattsson96b}
                M.S. Mattsson, G.A. Niklasson and C.G. Granqvist,
	        {\it Phys. Rev.} {\bf B54}, 17884 (1996). 


\bibitem{cherry96}
		M. L. Cherry et al,
	        {\it Phys. Rev. } {\bf C53}, 1532 (1996).

\bibitem{du96}
		R. Du and H. A. Stone,
	        {\it Phys. Rev. } {\bf E53}, 1994 (1996).

\bibitem{balazs96}
		L. Balazs, V. Fleury, F. Duclos, and A. Van Herpen,
	        {\it Phys. Rev. } {\bf E54}, 599 (1996).

\bibitem{muller96} 
                H.-P. Muller, R. Kimmich and J. Weis,
	        {\it Phys. Rev.} {\bf E54}, 5278 (1996). 

\bibitem{bisang96}
                U. Bisang and J.H. Bilgram 
	        {\it Phys. Rev.} {\bf E54}, 5309 (1996). 


\bibitem{mallamace96}
	        F. Mallamace, N. Micali, S. Trusso, L. M. Scolaro,
                A. Romeo, A. Terracina, and R. F. Pasternack, 
	        {\it Phys. Rev. Lett} {\bf 76}, 4741 (1996).

\bibitem{pignon96}
	        F. Pignon, J. M. Piau, and A. Magnin,
	        {\it Phys. Rev. Lett} {\bf 76}, 4857 (1996).


% ***** This is the end of the list of references for the Table ******


\bibitem{linear-log10}
{Conventionally, log of base 10 is used for both horizontal and vertical axes
  and the width of the linear range is given in decades. In cases that a
  different base was used we converted it to base 10. It is also important to
  realize that the property presented in the horizontal axis should have a
  linear dimension (such as length, time, etc.) as fractal dimension relates
  some generalized volume to a linear stick size. If the feature presented has
  dimensions of area, for example, the apparent width of the linear range will
  double. In a number of cases we had to correct the width measurements to
  avoid such effects.}

\bibitem{Dexceptions}
The only paper reporting a dimension much larger than 3 is
Ref.
\cite{komori94} 
where the fractal dimension of a strange attractor in 
turbulent plasma was measured.


\bibitem{Hentschel83}
H.G.E. Hentschel and I. Procaccia,
{\it Physica} {\bf D8}, 435 (1983).



\bibitem{Halsey86}
T.C. Halsey, M.H. Jensen, L.P. Kadanoff, I. Procaccia and B. Shraiman,
{it  Phys. Rev. A} {\bf 33}, 1141 (1986). 


\bibitem{Potschke91}
G. Potschke, J. Schroder, C. Gunther, R.Q. Hwang and R.J. Behm,
{\it Surf. Sci.} {\bf 251}, 592 (1991).


\bibitem{Bott92}
M. Bott, T. Michely and G. Comsa,
{\it Surf. Sci.} {\bf 272}, 161 (1992).

\bibitem{Michely93}
T. Michely, M. Hohage, M. Bott and G. Comsa,
{\it Phys. Rev. Lett.} {\bf 70}, 3943 (1993).


\bibitem{Stoyanov81}
S. Stoyanov and D. Kashchiev,
{\it Current Topics in Material Science}
{\bf 7}, 70 (1981).

\bibitem{Venables84}
J. A. Venables, G. D. T. Spiller, and M. Hanbucken,
{\it Rep. Prog. Phys.} {\bf 47}, 399 (1984).

\bibitem{Villain92}
J. Villain, A. Pimpinelli, and D. Wolf,
{\it Comments Cond. matt. Phys.} {\bf 16}, 1 (1992);
J. Villain, A. Pimpinelli, L. Tang, and D. Wolf,
{\it J. Phys. I France} {\bf 2}, 2107 (1992).

\bibitem{Bartelt92}
M.C. Bartelt and J.W. Evans,
{\it Phys. Rev. B} {\bf 46}, 12675 (1992);
M.C. Bartelt, M. C. Tringides and J.W. Evans,
{\it Phys. Rev. B} {\bf 47}, 13891 (1993).

\bibitem{Tang93}
L. Tang,
{\it J. Phys. I France } {\bf 3}, 935 (1993).

\bibitem{Biham95}
O. Biham, G.T. Barkema, and M. Breeman,
{\it Surf. Sci.} {\bf 324}, 47 (1995).


\bibitem{Ratsch94} 
C. Ratsch, A. Zangwill, P. Smilauer 
and D. D. Vvedensky, 
{\it Phys. Rev. Lett.} {\bf 72}, 3194 (1994);
C. Ratsch, P. Smilauer, A. Zangwill and D.D Vvedensky, 
{\it Surf. Sci.} {\bf 329}, L599 (1995).

\bibitem{Zhang94}
Z. Zhang, X. Chen and M.G. Lagally,
{\it Phys. Rev. Lett.} {\bf 73} 1829 (1994).

\bibitem{Jacobsen94}
J. Jacobsen, K.W. Jacobsen, P. Stoltze and J. K. Norskov,
{\it Phys. Rev. Lett.} {\bf }  (1994).

\bibitem{Barkema94}
G.T. Barkema, O. Biham, M. Breeman, D.O.Boerma,
and G. Vidali.
{\it Surf. Sci. Lett.} {\bf 306}, L569 (1994).

\bibitem{Schroeder95}
M. Schroeder and D.E. Wolf, 
{\it Phys. Rev. Lett.} {\bf 74}, 2062 (1995);
D. E. Wolf in {\it Scale invariance, Interfaces, and 
Non-Equilibrium Dynamics},
edited by M. Droz, A. J. McKane, J. Vannimenus 
and D. E. Wolf, NATO-ASI Series
(Plenum, New York, 1994).


\bibitem{Amar95}
J. G. Amar and F. Family,
{\it Phys. Rev. Lett.} {\bf 74}, 2066 (1995);
{\it Thin Solid Films} {\bf 272}, 208 (1996);
F. Family and J.G. Amar, 
{\it Mater. Sci. Eng. B} {\bf 30}, 149 (1995).

\bibitem{Bales95}
G.S. Bales and D.C. Chrzan, 
{\it Phys. Rev. Lett.} {\bf 74}, 4879 (1995).

\bibitem{Linderoth96}
T.R. Linderoth, J.J. Mortensen, K.W. Jacobsen, 
E. Laegsgaard, I Stensgaard and F. Besenbacher,
{\it Phys. Rev. Lett.} {\bf 77} 87 (1996).



\bibitem{Furman97}
I. Furman and O. Biham,
{\it Phys. Rev.} {\bf B55}, 7917 (1997).


\bibitem{third}
In the present analysis we focus on the case 
$i^{\ast}=1$ and
$\gamma=1/3$. 
In systems for which $\gamma \neq 1/3$ 
there are unstable or mobile islands of size 
$s \ge 2$.
Such systems exhibit mobility along island edges or 
detachment moves that modify
the morphology from fractal to more compact
even at the time-scale of 
single atom hopping,
and therefore are not relevant for our considerations. 

\bibitem{manipulation}
In principle, one can vary the deposition rate
during the growth process. 
This is typically used to increase the number 
of nucleation sites which is helpful for 
epitaxial growth
\cite{Rosenfeld93}.
This is achieved by starting with a high
deposition rate and gradually
reducing it as the coverage increases.
The large number of islands nucleated in the early
stages are stable and keep aggregating more atoms
in spite of the reduced deposition rate.
For the purpose of growing larger DLA-like islands
in a shorter time one may want to use a slow
deposition rate in the early stages and increase
it gradually. 
However, in this case new islands will continue to
nucleate and the low island density of the initial
low deposition rate will not be maintained.

\bibitem{Rosenfeld93}
G. Rosenfeld, R. Servaty, C. Teichert, B. Poelsma and G. Comsa,
{\it Phys. Rev. Lett.} {\bf 71}, 895 (1993).

\bibitem{Garey79}
M.R. Garey and D.S. Johnson,
{\it Computers and intractability: a guide to the theory of
NP-completeness} (Freeman, New York, 1979).


\bibitem{Stumpf94}
R. Stumpf and M. Scheffler, 
{\it Phys. Rev. Lett.} {\bf 72}, 254 (1994).

\bibitem{Stumpf96}
R. Stumpf and M. Scheffler, 
{\it Phys. Rev.} {\bf B53}, 4958 (1996).


\bibitem{Ayrault74}
G. Ayrault and G. Ehrlich,
{\it J. Chem. Phys.} {\bf 60}, 281 (1974).

\bibitem{Basset78}
D.W. Basset and P.R. Webber,
{\it Surf. Sci.} {\bf 70}, 520 (1978).

\bibitem{Kellogg91}
D.W. Basset and P.R. Webber,
{\it Surf. Sci.} {\bf 246}, 31 (1991).


\bibitem{Hamburger96a}
{D.A. Hamburger, O. Malcai, O. Biham and D. Avnir}, {in ``Fractals and Chaos in
  Chemical Engineering'', World Scientific, Singapore (1997).}

\bibitem{Hamburger96b}
{D.A. Hamburger, O. Biham and D. Avnir}, Phys. Rev. E {\bf 53},  3342  (1996).

\bibitem{Lidar97}
D.A. Lidar (Hamburger), O. Biham and D. Avnir,
{\it J. Chem. Phys.}, in press (1997).

\bibitem{Avnir97}
D. Avnir, O. Biham, D.A. Lidar (Hamburger) and O. Malcai,
in
{\it Fractal Frontiers}, Edited by M.M. Novak and T.G. Dewey
(World Scientific, Singapore, 1997), p. 199.


\end{thebibliography}
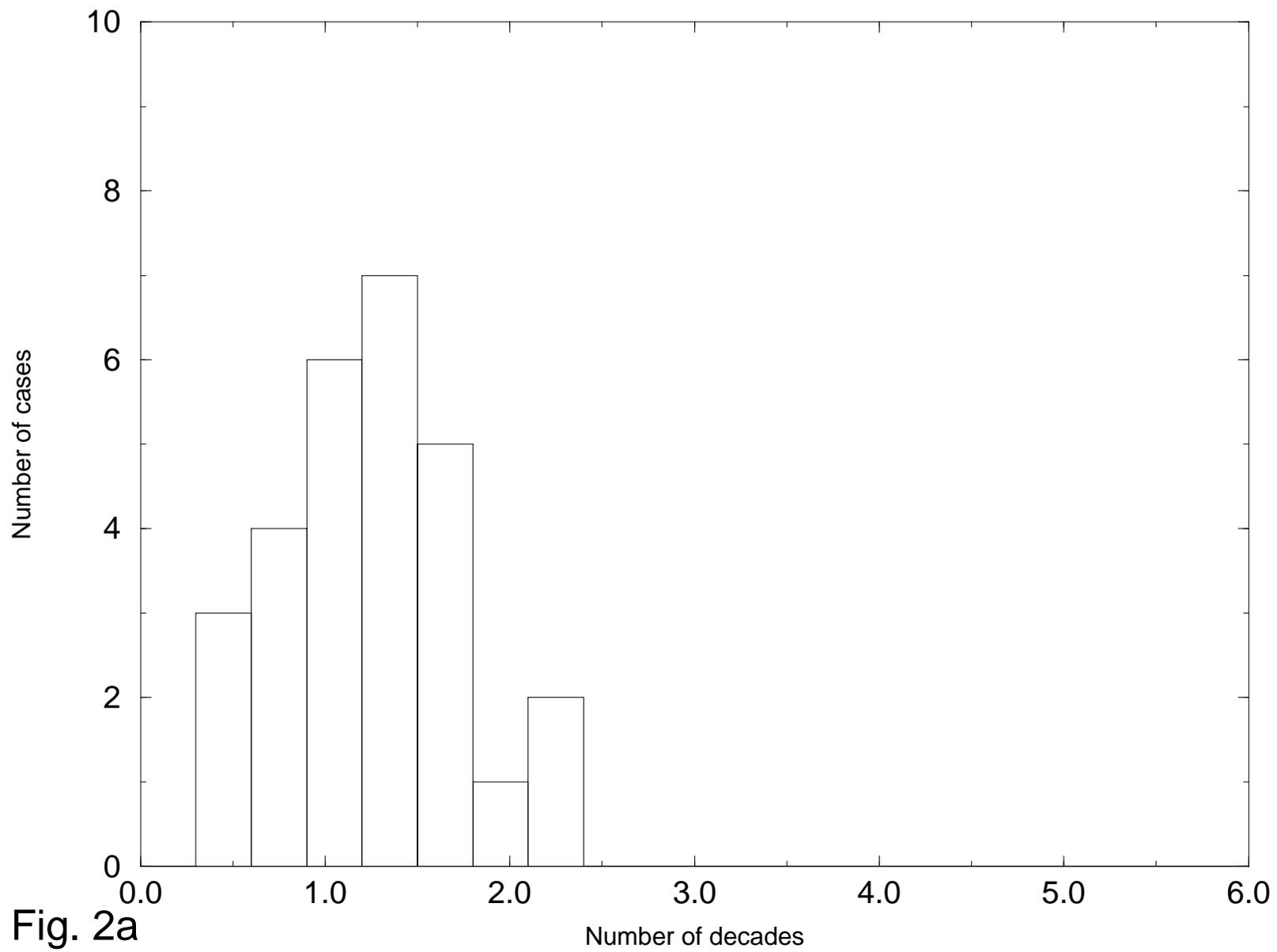
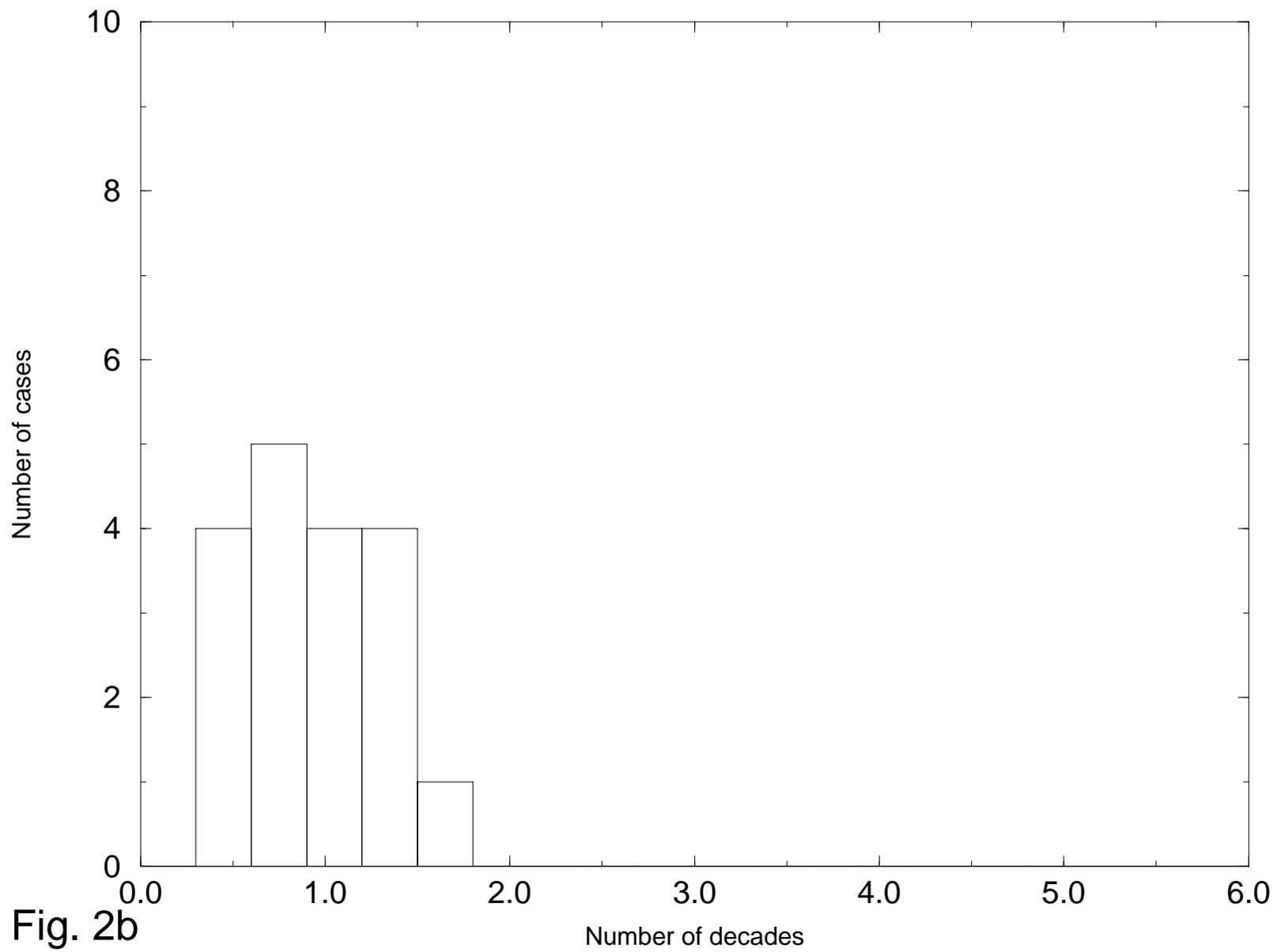
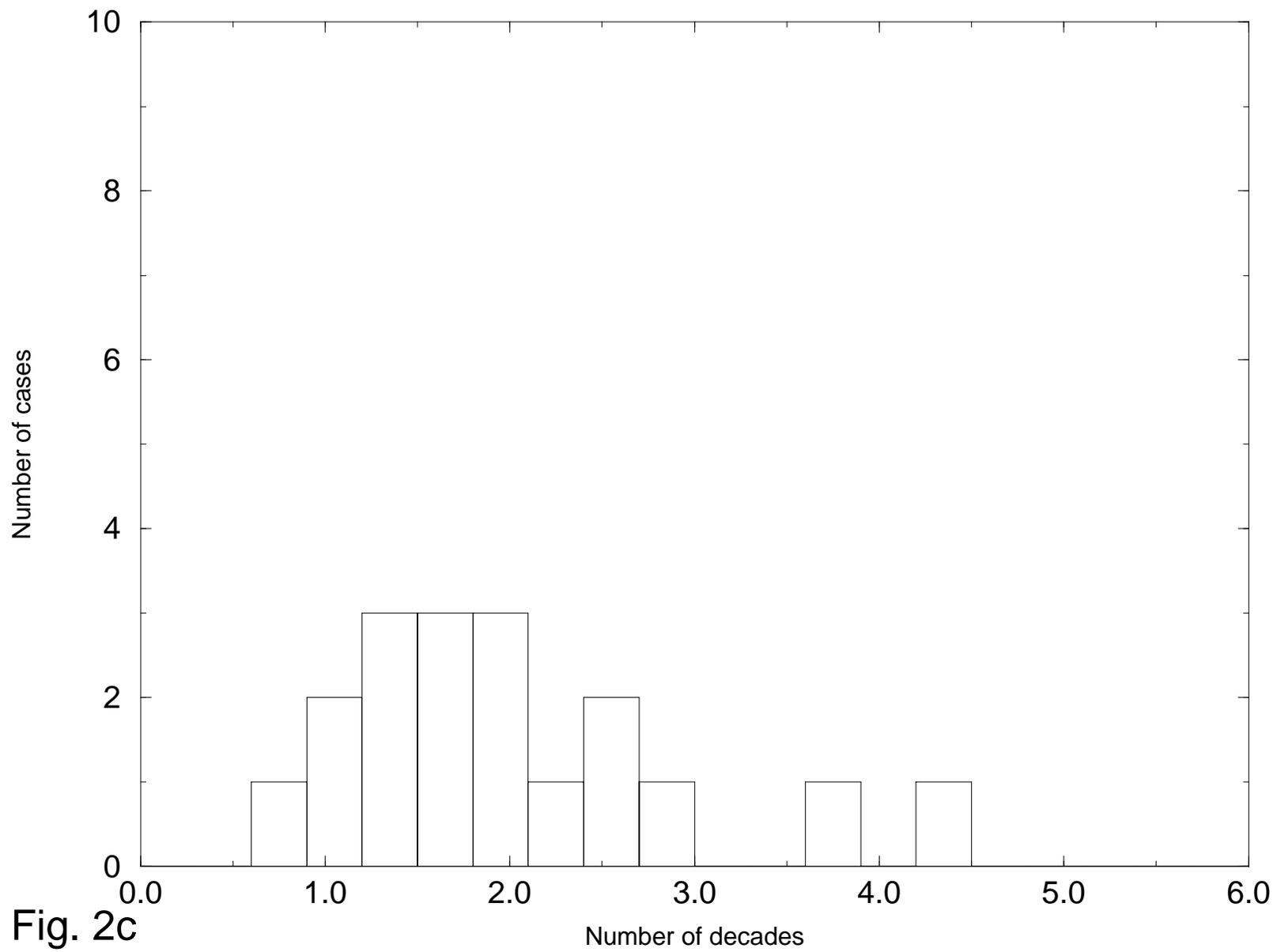
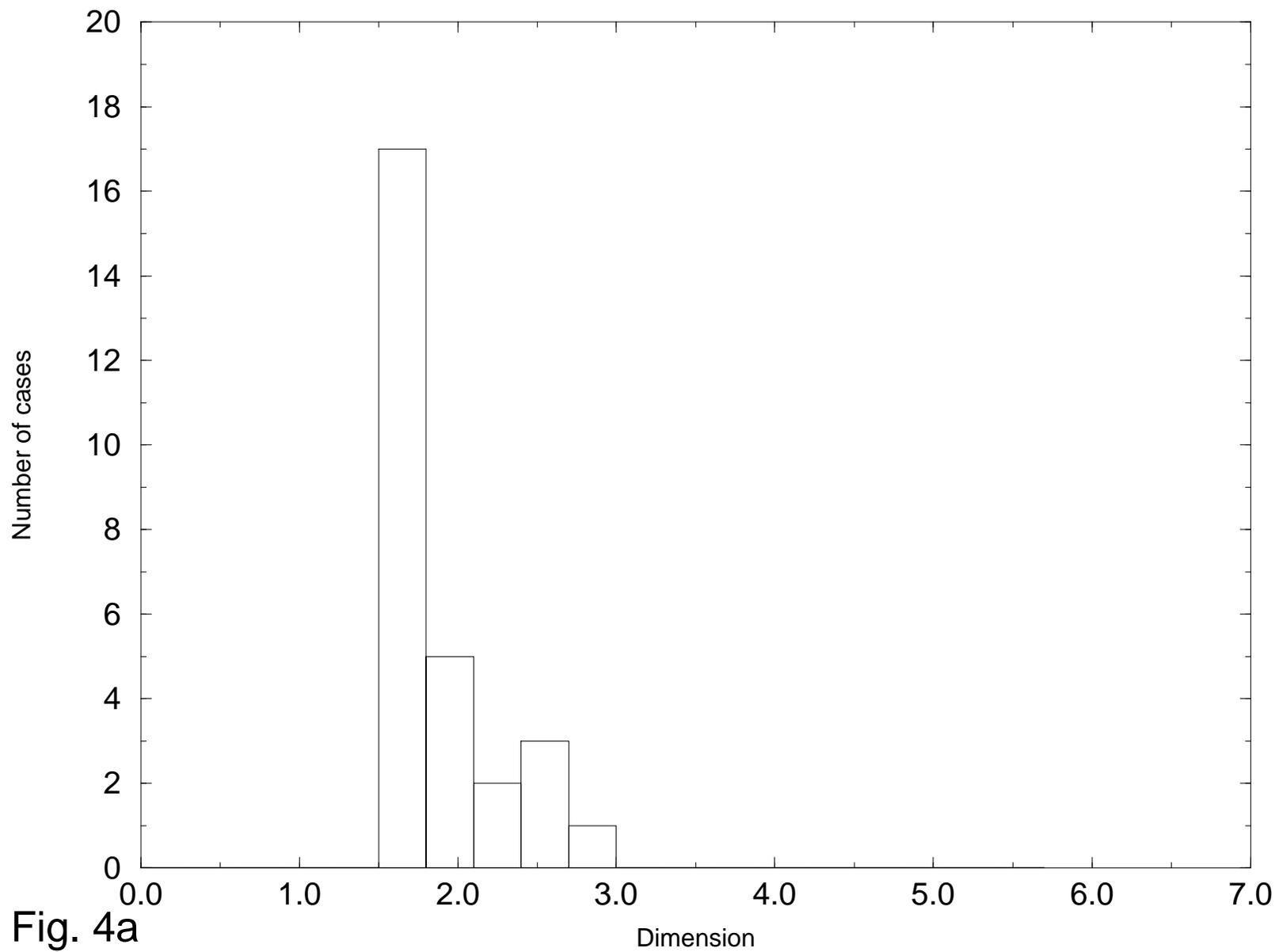
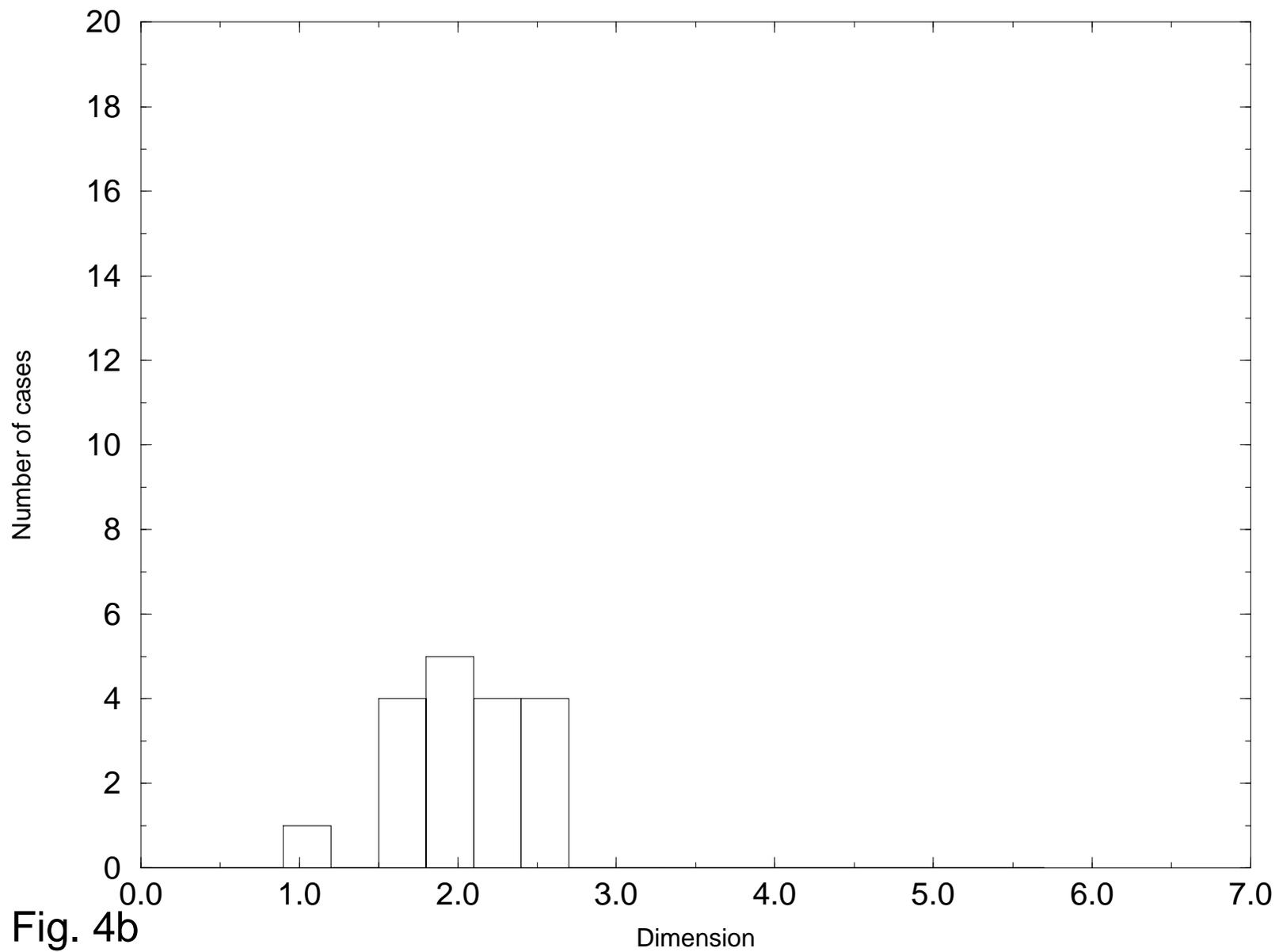
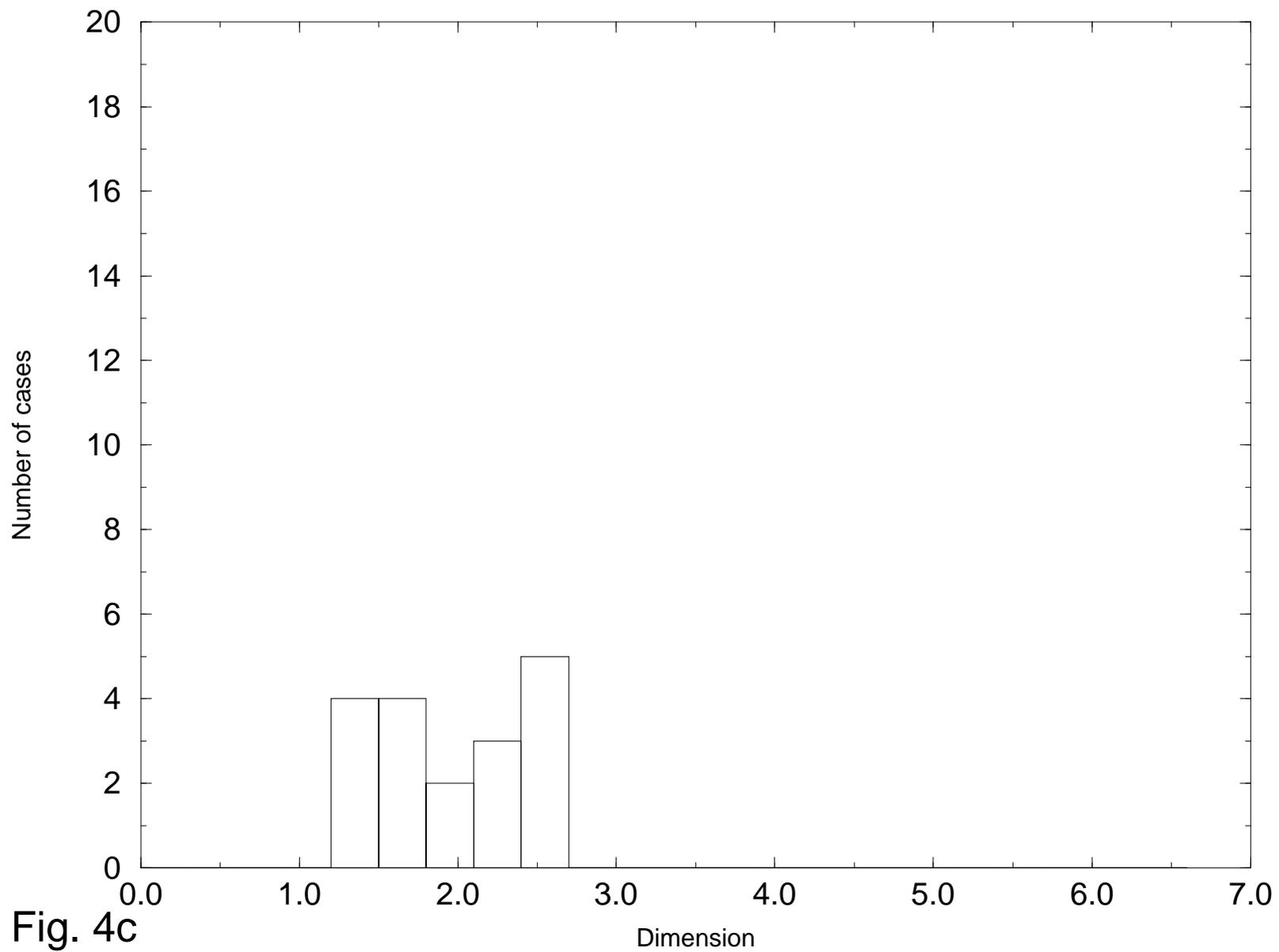
\end{document}